\documentclass[a4paper,12pt]{article}
\usepackage[utf8]{inputenc}
\usepackage[T1]{fontenc}
\usepackage[english]{babel}
\usepackage{graphicx}
\usepackage{mathtext}
\usepackage{indentfirst}
\usepackage{amsmath,amsfonts,braket,fp,bbm,tikz-cd,authblk,hyperref}
\usepackage{amssymb}
\usepackage{amsmath}
\usepackage{cases}
\usepackage{slashed}
\usepackage{euscript}
\usepackage{bbold}
\usepackage{bm,paralist,xspace,url,relsize}
\usepackage{color,xcolor}
\usepackage{hyperref}
\usepackage{textcomp}
\usepackage{appendix}
\usepackage{titlesec}
\usepackage{blindtext}
\usepackage{cmap} % поиск в PDF
\usepackage[left=1.5cm,right=1.5cm,top=2cm,bottom=2cm,bindingoffset=0cm]{geometry}
\PassOptionsToPackage{unicode}{hyperref}
\PassOptionsToPackage{naturalnames}{hyperref}
\begin{document}
 \setcounter{section}{0}

\renewcommand{\contentsname}{Contents}

%\begin{titlepage}

%\title{Non--stationary quantum fields in strong scalar wave background}

%\end{titlepage}

\author[1,2]{E.T.Akhmedov}
\author[3]{O.Diatlyk}
\author[1,4]{A.G. Semenov}
\affil[1]{Moscow Institute of Physics and Technology, 141700, Dolgoprudny, Russia}
\affil[2]{Institute for Theoretical and Experimental Physics, 117218, Moscow, Russia}
\affil[3]{National Research University Higher School of Economics, 101000 Moscow, Russia}
\affil[4]{I.E. Tamm Department of Theoretical Physics, P.N. Lebedev Physical Institute, 119991 Moscow, Russia}

\title{\textcolor{black}{Out of equilibrium two--dimensional Yukawa theory in a strong scalar wave background}}

%\begin{document}

\numberwithin{equation}{section}

\maketitle

\begin{abstract}

We consider 2D Yukawa theory in the strong scalar wave background. We use operator and functional formalisms. In the latter the Schwinger--Keldysh diagrammatic technique is used to calculate retarded, advanced and Keldysh propagators. We use simplest states in the two formalisms in question, which appear to be different from each other. As the result two Keldysh propagators found in different formalisms do not coincide, while the retarded and advanced ones do coincide. We use these propagators to calculate physical quantities. Such as the fermion stress energy flux and the scalar current. The latter one is necessary to know to address the backreaction problem. It happens that while in the functional formalism (for the corresponding simplest state) we find zero fermion flux, in the operator formalism (for the corresponding simplest state) the flux is not zero and is proportional to a Schwarzian derivative. Meanwhile the scalar current is the same in both formalisms, if the background field is large and slowly changing.

\end{abstract}
%\newpage

%///////////////////////////////////////

%\centerline{\Large \bf Out of equilibrium two--dimensional Yukawa theory}
%\centerline{\Large \bf in a strong scalar wave background}

%\vspace{10mm}

%\centerline{E. T. ${\rm Akhmedov}^{1, 2}$, O. ${\rm Diatlyk}^{3}$ and A. ${\rm Semenov}^{4}$}

%\vspace{5mm}

%\centerline{Abstract:}

%We consider 2D Yukawa theory in the strong scalar wave background. We use operator and functional formalisms. In the latter the Schwinger--Keldysh diagrammatic technique is used to calculate retarded, advanced and Keldysh propagators. We use simplest states in the two formalisms in question, which appear to be different from each other. As the result two Keldysh propagators found in different formalisms do not coincide, while the retarded and advanced ones do coincide. We use these propagators to calculate physical quantities. Such as the fermion stress energy flux and the scalar current. The latter one is necessary to know to address the backreaction problem. It happens that while in the functional formalism (and the corresponding simplest state) we find zero fermion flux, in the operator formalism (and the corresponding simplest state) the flux is not zero. Meanwhile the scalar current is the same in both formalisms, if the background field is large and slowly changing.

\newpage

\tableofcontents

\newpage

\section{Introduction}

There is the long standing backreaction problem in strong background fields (see e.g. old textbooks \cite{Birrell:1982ix}, \cite{GriMamMos} for the introduction). In this respect it would be nice to have a simple enough, but non--trivial example of an out of equilibrium QFT in a strong background field. One possible option is to consider QFT in strong electric or gravitational fields in 2D rather than in 4D. However, there the electromagnetic and gravitational fields are essentially non--dynamical, while to study the backreaction problem it is more appropriate to have dynamical fluctuations over the background field. Then an option is to study strong scalar field background in a 2D QFT.

Namely, we propose to consider Yukawa theory of fermions interacting with massless real scalar field in (1+1)-dimensional Minkowski spacetime with the action:
\begin{equation}
    \label{M:1}
   S[\psi, \bar{\psi},\phi]=  \int d^{2}x \bigg( \dfrac{1}{2}\partial_{\mu}\phi \partial^{\mu}\phi+ \bar{\psi}i \slashed{\partial}\psi-\lambda \phi\bar{\psi}\psi \bigg).
\end{equation}
The signature of the metric is (1,-1);
Gamma matrices have the form:

\begin{equation}
    \label{iqas1}
    \gamma^{0}=\left[ \begin{array}{cc}
   0 & 1 \\
    1 & 0 \\
  \end{array} \right],  \qquad  \gamma^{1}=\left[ \begin{array}{cc}
   0 & 1 \\
    -1 & 0 \\
  \end{array} \right].
\end{equation}
In the presence of classical background fields we split $\psi=\psi_{cl}+\psi_{q}$ and $\phi=\phi_{cl}+\phi_{q}$, where $\phi_{cl}$, $\psi_{cl}$ are solutions of classical equations of motion and $\phi_q$, $\psi_q$ are quantum fluctuations:

\begin{equation}
    \label{M:2}
    \begin{cases}
    \partial^{2}\phi_{cl} + \lambda \, \bar{\psi}_{cl} \, \psi_{cl} = 0 \\
    \big[ i \gamma^{\mu}\partial_{\mu} - \lambda \, \phi_{cl}\big]\, \psi_{cl} = 0 \ .
    \end{cases}
\end{equation}
In particular we consider the following scalar wave solution:

\begin{equation}
    \label{M:3}
    \lambda \, \phi_{cl}(t,x) = \Phi\bigg(\dfrac{t-x}{\sqrt{2}}\bigg) \quad {\rm and} \quad
    \psi_{cl}=0.
\end{equation}
Other options $\phi_{cl} = \alpha + \beta\, t$ and $\phi_{cl} = \alpha + \beta \, x$, for constant and real $\alpha$ and $\beta$, will be considered in a separate paper.

However, the background scalar fields seem to suffer from a number of disadvantages because do not share some of the relevant properties of the strong electric and gravitational fields. In fact, consider a point like relativistic particle in a scalar field. The simplest classical action in such a situation is as follows:

$$
S = - \int d\tau \,\left\{m^{\phantom{\frac12}} + \,\, \lambda \, \phi[x(\tau)]\right\},
$$
where $\tau$ is the proper time, $\phi(x)$ is the scalar field, $x(\tau)$ is the world--line of the particle, $\lambda$ is its ``charge'' with respect to the scalar field and $m$ is its mass.

The equation of motion that follows from this action is:

\begin{equation}\label{claseq}
\left[m + \lambda \, \phi\right] \, \ddot{x}_\mu = \lambda \, \left[\dot{x}_\nu \,\partial_\mu \phi - \partial_\nu \phi \, \dot{x}_\mu\right] \, \dot{x}^\nu, \quad \mu, \nu = \overline{0,1} \, ,
\end{equation}
and $\dot{x}_\mu \, \dot{x}^\mu = 1$. One can show that for the examples of the background scalar fields that have been listed around eq. (\ref{M:3}) the analytically continued eq. (\ref{claseq}) does not have Euclidian world--line instanton solutions. Furthermore, in this paper we will also see that the fermionic effective action in such background fields does not have an imaginary contribution and it is analytical on the cutted complex plane of the background field. All this means that there is no particle tunneling in background scalar fields under consideration, unlike the strong electric and gravitational fields. However, as we show in this paper the situation is not that trivial and in this simple Yukawa theory there are interesting effects, which are related to the particle creation.

To begin with, in this paper we neglect quantum fluctuations of the scalar field $\phi_q$. (Potentially highly important, as we explain in the concluding section, loop effects due to quantum scalar fluctuations will be considered in a separate paper.) The action in such a case simplifies to

\begin{equation}
    \label{eq:2}
   S[\psi, \bar{\psi}] = \int d^2x \bigg( \bar{\psi}(x,t)i \slashed{\partial}\psi(x,t)-\Phi\bigg(\dfrac{t-x}{\sqrt{2}}\bigg)\bar{\psi}(x,t)\psi(x,t) \bigg).
\end{equation}
We assume that the potential $\Phi(v)$ vanishes at $v=\pm \infty$.

This theory is simple, since it is Gaussian, but is not trivial, because its action and Hamiltonian contain explicit time dependence. But still, we expect that it is in a sense exactly solvable, if the field $\Phi$ is not dynamical.

The aim of this paper is to achieve a better understanding of the non--equilibrium time evolution of quantum theory in strong background fields. To do that we consider two different approaches to the same model: functional and operator formalisms. Also we choose in these approaches different initial states, which are the simplest possibilities in the corresponding frameworks. The aim is not to simply test different formalisms, but rather to find how they can be matched to each other and to explain why the discrepancies in the answers for physical observables do appear.

In both cases we calculate the retarded, advanced and Keldysh propagators, which are the building blocks for physical observables in the theory. We find that the retarded and advanced propagators calculated in both formalisms are the same, which is appropriate, as is explained in the main body of the paper. However, the Keldysh propagators are different, because they are sensitive to the states of the theory. Difference in the Keldysh propagators leads to the differences in the physical observables. At the end of the paper we discuss a relation between the Keldysh propagators found in different settings.

What sort of observables are we interested in?
We calculate the expectation value of the fermionic stress energy tensor and of the scalar current, $\bar{\psi}\psi$, operator. The reason to consider the latter is the following. From the Hamiltonian of the theory \eqref{M:1} one obtains the operator equation:

\begin{equation}
    \label{M:8}
    \partial^{2}\hat{\phi}+\lambda\hat{\bar{\psi}}\hat{\psi}=0 \ ,
\end{equation}
which reproduces one of the classical equations of motion \eqref{M:2}. To solve this equation iteratively we take expectation values of its both sides and use the method of successive approximations. At the leading order in the expansion over the quantum fluctuations $\phi_q$ and $\psi_q$ we reproduce the first equation in \eqref{M:2} with $\psi_{cl} = 0$. Then, we put $\phi=\phi_{cl}$ to the second equation in \eqref{M:2} and solve for fermionic field. After that we can solve averaged eq. (\ref{M:8}) for $\langle \phi\rangle$ when $\langle \bar{\psi}\psi\rangle$ is calculated in the background of $\phi=\phi_{cl}$.
Thus, as follows from \eqref{M:8}, the expectation value $\langle \hat{\bar{\psi}} \hat{\psi}\rangle$ calculated in the background of $\phi_{cl}$ serves as a response of quantum fluctuations of fermionic field on the scalar background field.

In this paper we show that for those values of $\phi_{cl}$ which are big and slowly varying the expectation value of the scalar current, $\bar{\psi} \, \psi$, is the same in both formalisms, that have been briefly described above. As we explain below there is some sort of universality in the dependence of scalar current on the background field for such states that lead to Green functions with the proper Hadamard behaviour.

However, in the expectation values of the stress energy tensor we find a disagreement in the two formalisms. In particular, in one formalism we find zero fermion flux, while in the other it is not zero. As we explain the disagreement comes from the fact that different states are used in the calculations in the two different formalisms.

The paper is organized as follows. In the section 2 we briefly describe the uses of the Schwinger--Kledysh diagrammatic technique in the functional formalism. We derive the Dyson--Schwinger equations for the raterded, advanced and Keldysh propagators and solve them. Then, we use the resulting Keldyh propagator to calculate the physical observables.

In the section 3 we use the operator formalism to quantize the Yukawa theory in the strong scalar wave background. We find the exact fermionic basis of modes in this background and then use it to find the retarded, advanced and Keldysh propagators. The retarded and advanced propagators found in this formalism coincide with those found in the functional one. Kledysh propagators do not coincide. Again we use the new Keldysh propagator to find the physical observables.

In the section 4 we give an explanation where the disagreement between the two Keldysh propagators comes from. We derive an equation relating two Keldysh propagators, but cannot solve it yet. In the section 5 we describe peculiarities of the two found in the paper Keldysh propagators. We conclude in the section 6. To make the paper self contained and to simplify the presentation in its main body we put most of the calculations into the Appendix sections.

\section{Functional formalism and the corresponding simplest state}\label{1}

In non--stationary situations the quantities to consider are the correlation functions (see e.g. \cite{LL}, \cite{Kamenev} and \cite{Arseev}):

\begin{equation}\label{Ottt}
\langle O(t_1, \dots, t_n)\rangle = \left\langle st\left| U^+ \, T[O(t_1, \dots, t_n) \, U] \right|st \right\rangle
\end{equation}
rather than amplitudes:

\begin{equation}\label{Ottt1}
A = \frac{\left\langle out\left|T[O(t_1, \dots, t_n) \, U] \right|in \right\rangle}{\left\langle out\left| U \right|in \right\rangle },
\end{equation}
at least because there are no asymptotic states.
The amplitudes are more appropriate to calculate in stationary situations (in the proper ground state).
Here $ O(t_1, \dots, t_n)$ is an operator in the theory under consideration in the interaction picture; $U = T\exp{i \, \int_{-\infty}^{+\infty} dt' \, H_{int}(t')}$ is the evolution operator, where $H_{int}(t)$ is the non--linear part of the full Hamiltonian in the interacting picture, $\left|st \right\rangle$ is a state out of equilibrium, while $\left|in \right\rangle$ ($\left|out \right\rangle$) is the true ground state (rotated by a phase) of the normal ordered free Hamiltonian, if such a state does exist\footnote{Turning on and switching off of the interaction term $H_{int}$ at past and future infinities is usually assumed in the stationary situations. Note that (\ref{Ottt}) reduces to (\ref{Ottt1}) if there the quantum average is taken over the true ground state $|st \rangle \to | in \rangle$ and if $|\langle out| U |in \rangle| = 1$ under the adiabatic turning on and switching off $H_{int}$. Meanwhile the direct calculation of loop corrections to (\ref{Ottt1}) in a non--stationary situation leads to loop infrared divergences that cannot be cancelled out \cite{Akhmedov:2009be}, \cite{Akhmedov:2009vh}, \cite{Akhmedov:2008pu}.}.

In eq. (\ref{Ottt1}) all the expressions are time--ordered and, hence, one can apply the Feynman diagrammatic technique. At the same time, if one converts (\ref{Ottt}) into the functional integral form \cite{Kamenev}, then there are two copies of the action, which appear in the exponent under the integral. One is coming from $U$ and the other --- from $U^+$. That is how one obtains the so called Schwinger--Keldysh time contour $\mathcal{C}$, which goes forward from past to future infinity and then back.  For convenience we denote fields on the forward branch of the contour $\mathcal{C}$ as $\psi(t_{+},x) \equiv \psi_{+}(t,x)$ and the ones on the backward branch as $\psi(t_{-},x) \equiv \psi_{-}(t,x)$  \cite{Kamenev}. Then the action in the exponent under the functional integral can be rewritten in terms of these fields as:

\begin{eqnarray}
    \label{eq:i2}
     S[\psi,\bar{\psi}]=\int_{-\infty}^{+\infty}dt \int dx \bigg[ \bar{\psi}_{+}(x,t)i \slashed{\partial}\psi_{+}(x,t)-\Phi\bigg(\dfrac{t-x}{\sqrt{2}}\bigg)\bar{\psi}_{+}(x,t)\psi_{+}(x,t)- \nonumber \\
     -\bigg(\bar{\psi}_{-}(x,t)i \slashed{\partial}\psi_{-}(x,t)-\Phi\bigg(\dfrac{t-x}{\sqrt{2}}\bigg)\bar{\psi}_{-}(x,t)\psi_{-}(x,t)\bigg)\bigg],
\end{eqnarray}
where the relative minus sign comes from the reversed direction of the time integration on the backward part of the contour, because $U^+$ contains the complex conjugate exponent with respect to $U$.

Then, the propagators are defined as follows:

\begin{equation}
    \label{eq:i1}
    iG(t,x;t^{\prime},x^{\prime}) \equiv \int D \psi D \bar{\psi} \ \psi(t,x)\bar{\psi}(t^{\prime},x^{\prime}) \, e^{iS[\psi, \bar{\psi}]} = \braket{\psi(t,x)\bar{\psi}(t^{\prime},x^{\prime})}.
\end{equation}
In terms of $\psi_{+}$ and $\psi_{-}$ we have the following matrix of propagators:

\begin{equation}
    \label{eq:3}
   \hat{G}(t,x;t^{\prime},x^{\prime})=
  \left[ {\begin{array}{cc}
   -i\braket{T\psi_{+}(t,x)\bar{\psi}_{+}(t^{\prime},x^{\prime})} & i\braket{\bar{\psi}_{-}(t^{\prime},x^{\prime})\psi_{+}(t,x)} \\
   -i\braket{\psi_{-}(t,x)\bar{\psi}_{+}(t^{\prime},x^{\prime})} & -i\braket{\bar{T}\psi_{-}(t,x)\bar{\psi}_{-
  }(t^{\prime},x^{\prime})} \\
  \end{array} } \right] \equiv \left[ {\begin{array}{cc}
   G_{++} & G_{<} \\
    G_{>} & G_{--} \\
  \end{array} } \right],
\end{equation}
 where\footnote{It is worth to mention that $G_{>}$, $G_{<}$ and $G_{\pm \pm}$ are $2\times2$ matrixes in spinor indexes. So, $G$ itself is a block matrix of $2\times2$ matrices.}
 \begin{eqnarray}
     \label{eq:4}
     G_{++}(t,x;t^{\prime},x^{\prime}) = \theta(t-t^{\prime}) \, G_{>}(t,x;t^{\prime},x^{\prime}) + \theta(t^{\prime}-t) \, G_{<}(t,x;t^{\prime},x^{\prime}), \nonumber \\
     G_{--}(t,x;t^{\prime},x^{\prime}) = \theta(t-t^{\prime}) \, G_{<}(t,x;t^{\prime},x^{\prime}) + \theta(t^{\prime}-t) \, G_{>}(t,x;t^{\prime},x^{\prime}) \ .
 \end{eqnarray}
From this definition it is clear, that:

\begin{equation}
    \label{eq:6}
    G_{++}+G_{--}=G_{>}+G_{<} \ .
\end{equation}
The structure of the action \eqref{eq:i2} and of the propagator \eqref{eq:i1} may lead to the conclusion that non-diagonal components of the propagator matrix \eqref{eq:3} must vanish. In that sense, functional integral representation of the theory is a bit misleading: it does not contain information about the initial state of the theory which makes $\psi_{+}$ and $\psi_{-}$ fields correlated (see e.g. \cite{Kamenev}). And if one works in the functional integral formalism it seems to be unclear where is the information about initial state of the fermions is hidden. Then it seems that one should always keep in mind operator formalism, from which the propagators can be derived using initial density matrix, as it is done in Appendix A.

It is possible to make the presence of the initial state apparent in the functional formalism by doing the so called Keldysh rotation \cite{LL}, \cite{Kamenev}. In fact, if we introduce the new pair of fields:

\begin{equation}
    \label{eq:7}
\begin{cases}
\Psi_{1}=\dfrac{1}{\sqrt{2}}\bigg(\psi_{+}+\psi_{-}\bigg) \\ \Psi_{2}=\dfrac{1}{\sqrt{2}}\bigg(\psi_{+}-\psi_{-}\bigg)
\end{cases}
\quad \quad \text{and} \qquad \qquad \begin{cases}
\Bar{\Psi}_{1}=\dfrac{1}{\sqrt{2}}\bigg(\bar{\psi}_{+}-\bar{\psi}_{-}\bigg) \\ \Bar{\Psi}_{2}=\dfrac{1}{\sqrt{2}}\bigg(\bar{\psi}_{+}+\bar{\psi}_{-}\bigg),
\end{cases}
\end{equation}
the action (\ref{eq:i2}) acquires the following form:

\begin{equation}
    \label{eq:8}
    S=\int_{-\infty}^{+\infty} dt dx \bigg[ \bar{\Psi}(x,t)i \slashed{\partial}\Psi(x,t)-\bar{\Psi}(x,t)\hat{\Phi}(t,x)\Psi(x,t) \bigg].
\end{equation}
where
\begin{equation}
    \label{eq:9}
    \Psi=\begin{bmatrix}\Psi_{1} \\ \Psi_{2} \end{bmatrix},
\quad \quad \Bar{\Psi}=\begin{bmatrix}\Bar{\Psi}_{1} \\ \Bar{\Psi}_{2} \end{bmatrix}
\quad \text{and} \quad  \hat{\Phi}(t,x)=\left[ {\begin{array}{cc}
   \Phi\bigg(\dfrac{t-x}{\sqrt{2}}\bigg)& 0 \\
    0 & \Phi\bigg(\dfrac{t-x}{\sqrt{2}}\bigg) \\
  \end{array} } \right].
\end{equation}
After such a rotation propagator matrix transforms into the triangular form:

\begin{equation}
    \label{eq:10}
   \hat{G}(t,x;t^{\prime},x^{\prime})=
  \left[ {\begin{array}{cc}
   -i\braket{\Psi_{1}(t,x)\bar{\Psi}_{1}(t^{\prime},x^{\prime})} & -i\braket{\Psi_{1}(t,x)\bar{\Psi}_{2}(t^{\prime},x^{\prime})} \\
   0 & -i\braket{\Psi_{2}(t,x)\bar{\Psi}_{2
  }(t^{\prime},x^{\prime})} \\
  \end{array} } \right]=\left[ {\begin{array}{cc}
   G^{R} & G^{K} \\
    0 & G^{A} \\
  \end{array} } \right],
\end{equation}
where

\begin{eqnarray}
\label{eq:i5}
    G^{R}(t,x;t^{\prime},x^{\prime})=\theta(t-t^{\prime})\bigg(G^{>}(t,x;t^{\prime},x^{\prime})-G^{<}(t,x;t^{\prime},x^{\prime})\bigg), \nonumber \\
    G^{A}(t,x;t^{\prime},x^{\prime})=\theta(t^{\prime}-t)\bigg(G^{<}(t,x;t^{\prime},x^{\prime})-G^{>}(t,x;t^{\prime},x^{\prime})\bigg), \nonumber \\
   G^{K}(t,x;t^{\prime},x^{\prime})=G^{>}(t,x;t^{\prime},x^{\prime})+G^{<}(t,x;t^{\prime},x^{\prime}).
\end{eqnarray}
The tree--level retarded and advanced Green functions (first two equations in \eqref{eq:i5}) do not dependent on the initial state of the theory: they only carry information about causality and spectrum. That is because these propagators are proportional to the anti--commutator of $\psi$'s, which is the c--number. Whereas the Keldysh propagator (the last line in \eqref{eq:i5}) does contain information about the initial density matrix. That is exactly the piece of information one needs to correctly define the correlation functions in the functional integral formalism (\ref{eq:i1}) (see e.g. \cite{Kamenev} and \cite{Arseev}).

In this section we assume that initially, at $t=-\infty$, when $\Phi(t-x) = 0$, fermions are at the equilibrium and have thermal distribution with a temperature $T$. (However, at some point below, to simplify expressions and keep them in a physically tractable form, we will have to put $T=0$, i.e. to consider fermions at the ground state at past infinity.)  Then, as is explained in Appendix \ref{A}, at past infinity the time Fourier transformation of the Keldysh propagator is related to the retarded and advanced propagators in the following way\footnote{Equation \eqref{eq:i8} constitutes the statement of the so called fluctuation–dissipation
theorem (FDT) \cite{LL}, \cite{Kamenev}: it implies a rigid relation between the response functions and the correlation functions in equilibrium.}:

\begin{equation}
    \label{eq:i8}
     G_{0}^{K}(\epsilon,x,x^{\prime}) = F(\epsilon) \, \bigg[G_{0}^{R}(\epsilon,x,x^{\prime})-G_{0}^{A}(\epsilon,x,x^{\prime})\bigg],
\end{equation}
where the index $"0"$ means that we consider the propagators in the absence of the background field, i.e. when $\Phi=0$, and

\begin{equation}
    \label{eq;i1000}
    F(\epsilon) = 1 - 2 \, n_\epsilon = 1 - 2 \, \dfrac{1}{e^{\epsilon/T}+1} = \tanh{\dfrac{\epsilon}{2T}},
\end{equation}
defines the distribution function; $G_0^{K,R,A}(\epsilon,x,x^{\prime})$ are time Fourier transformations of the Keldysh, retarded and advanced propagators, correspondingly.

In all, the theory defined with the help of Keldysh rotation (\ref{eq:7}) is self consistent, i.e. after establishing the relation \eqref{eq:i8} in operator formalism we can forget about this formalism and work only in the functional approach, knowing that all information about the initial density matrix is encoded in \eqref{eq:i8}.

\subsection{Dyson equation and R-A junction}\label{2}

To find the exact propagator matrix we treat the $\bar{\Psi}\hat{\Phi}\Psi$ term in (\ref{eq:8}) as a perturbation and use the causality condition \cite{Kamenev}:

$$
G^{R/A}(t,t_{1}) \cdot ... \cdot G^{R/A}(t_{n},t)=0,
$$
which follows from the fact that the retarded and advanced propagators are proportional to the $\theta$--functions. (Physically that just means that exact $G^{R/A}$ keep their property of being retarded and advanced Green functions, correspondingly, as their tree--level counterparts.)
This way we obtain the Dyson--Schwinger equation for the matrix of exact propagators:

\begin{eqnarray}
    \label{eq:14}
    \hat{G}(t,x;t^{\prime},x^{\prime}) = \hat{G}_{0}(t,x;t^{\prime},x^{\prime}) + \int d\tau dy \, \hat{G}_{0}(t,x;\tau,y) \, \hat{\Phi}(\tau,y) \, \hat{G}(\tau,y;t^{\prime},x^{\prime}) \equiv \nonumber \\ \equiv \hat{G}_{0} + \hat{G}_{0} \circ \hat{\Phi} \circ \hat{G}.
\end{eqnarray}
Where $\hat{G}_0$ is the matrix of propagators for $\Phi(v) = 0$, and $\hat{G}$ is the matrix of exact propagators.

In components this equation can be written as:

\begin{equation}
    \label{eq:16}
    \begin{cases}
G^{R}=G^{R}_{0}+G^{R}_{0} \circ \Phi \circ G^{R} \ , \\
G^{A}=G^{A}_{0}+G^{A}_{0} \circ \Phi \circ G^{A} \ , \\
G^{K}=G^{K}_{0}+G^{K}_{0} \circ \Phi \circ G^{A}+G^{R}_{0} \circ \Phi \circ G^{K} \ .
\end{cases}
\end{equation}
One can see that because the matrix $\hat{\Phi}(t,x)$  is diagonal the equations for $G^{R/A}$ are independent from each other. The derivation of the explicit form of $G^{K,R,A}_0$ can be found in Appendix A.

Since in the presence of $\Phi(v)$ there is no time translational invariance, in the exact propagators one has to perform the Fourier transformation in $t$ and $t'$ separately:

\begin{equation}
    \label{eq:125}
    G(\epsilon,x;\epsilon^{\prime},x^{\prime})\equiv \int dt dt^{\prime} \ G(t,x;t^{\prime},x^{\prime})e^{i\epsilon t-i\epsilon^{\prime} t^{\prime}}.
\end{equation}
At the same time when initially (at past infinity) fermions are at thermal equilibrium the Fourier transformation of the Keldysh propagator has the form \eqref{eq:i8} or we can rewrite it as:

\begin{equation}
    \label{eq:126}
    G^{K}_{0}(\epsilon,x;\epsilon^{\prime},x^{\prime})= G_{0}^{R}(\epsilon,x;\epsilon^{\prime},x^{\prime})F(\epsilon^{\prime})-F(\epsilon)G_{0}^{A}(\epsilon,x;\epsilon^{\prime},x^{\prime}) \ ,
\end{equation}
where $G^{K,R,A}_{0}(\epsilon,x;\epsilon^{\prime},x^{\prime}) \equiv G^{K,R,A}_{0}(\epsilon,x,x^{\prime})\delta(\epsilon-\epsilon^{\prime})$. That is because there is the time translational invariance in the case when $\Phi(v)=0$.

Making the inverse Fourier transformation of (\ref{eq:126}), we obtain:
\begin{equation}
    \label{eq:127}
    G^{K}_{0}(t,x;t^{\prime},x^{\prime})=\int d\tau G_{0}^{R}(t,x;\tau,x^{\prime})f(\tau-t^{\prime})-\int d\tau f(t-\tau)G_{0}^{A}(\tau,x;t^{\prime},x^{\prime})=G_{0}^{R} \circ f-f\circ G_{0}^{A} \ ,
\end{equation}
where the Fourier transformation of $F(\epsilon)$ is:

\begin{equation}
    \label{eq:52}
    f(\tau)=\int_{-\infty}^{+\infty} \dfrac{d\epsilon}{2\pi}e^{-i\epsilon \tau}\tanh\dfrac{\epsilon}{2T}=-\mathcal{P}\dfrac{iT}{\sinh{\big(\pi T \tau\big)}} \ ,
\end{equation}
as is explained at the end of Appendix \ref{A}.

With the use of the first two equations in (\ref{eq:16}) the last one there can be rewritten as:

\begin{equation}
    \label{eq:51}
    G^{K}(t,x;t^{\prime},x^{\prime})=\int d\tau \,  G^{R}(t,x;\tau,x^{\prime}) \, f(\tau-t^{\prime}) - \int d\tau \, f(t-\tau) \, G^{A}(\tau,x;t^{\prime},x^{\prime}) +
\end{equation}
$$+\int dy \int d\tau_{1} \int d\tau_{2} \, G^{R}(t,x;\tau_{1},y) \, G^{A}(\tau_{2},y;t^{\prime},x^{\prime}) \, \bigg[\Phi\bigg(\dfrac{\tau_2-y}{\sqrt{2}}\bigg) - \Phi\bigg(\dfrac{\tau_1-y}{\sqrt{2}}\bigg)\bigg] \, f(\tau_{1}-\tau_{2}),$$
i.e. we have expressed the exact Kledysh propagator via the exact retarded and advanced ones. For convenience, let us denote by $G^K_{th}$ sum of first two terms which reproduces the structure of the equilibrium Keldysh propagator \eqref{eq:127}. The third "anomalous" term is the so called R-A junction \cite{5}, \cite{6} which we denote as $G^K_{an}$. Note that the "anomalous" term vanishes in the case of constant potential $\Phi(v)$, while for non--trivial $\Phi(v)$ the theory is out of thermal equilibrium.

\subsection{Solution of the Dyson equation for the retarded and advanced propagators} \label{4}

Multiplying both sides of the first equation of the system \eqref{eq:16} by $\big[G_{0}^{R}\big]^{-1}$ we get:

\begin{equation}
    \label{eq:24}
    \bigg[\big[G_{0}^{R}\big]^{-1}-\Phi\bigg(\dfrac{t-x}{\sqrt{2}}\bigg)\bigg]\circ \, G^{R} = 1.
\end{equation}
Using that $\big[G_{0}^{R}\big]^{-1}=i\slashed{\partial}$, the last equation in components can be written as:

\begin{equation}
    \label{eq:25}
    \left[ {\begin{array}{cc}
   -\Phi\bigg(\dfrac{t-x}{\sqrt{2}}\bigg) & i\big(\partial_{t}+\partial_{x}\big) \\
    i\big(\partial_{t}-\partial_{x}\big) & -\Phi\bigg(\dfrac{t-x}{\sqrt{2}}\bigg) \\
  \end{array} } \right]
  \left[ {\begin{array}{cc}
   G^{R}_{11}(t,x;t^{\prime},x^{\prime})  & G^{R}_{12}(t,x;t^{\prime},x^{\prime}) \\
    G^{R}_{21}(t,x;t^{\prime},x^{\prime}) & G^{R}_{22}(t,x;t^{\prime},x^{\prime}) \\
  \end{array} } \right]=\delta(t-t^{\prime})\delta(x-x^{\prime})\hat{I},
\end{equation}
with the condition that

$$G^{R}(t,x;t^{\prime},x^{\prime})=0 \quad \quad \quad \text{if} \quad \quad \quad  t<t^{\prime}.$$
Since the potential $\Phi$ is a function  of $t-x$ only, it is convenient to work in the light-cone coordinates

\begin{equation}\label{LCcoord}
u=\dfrac{\big(t+x\big) }{\sqrt{2}} \quad {\rm and} \quad v=\dfrac{\big(t-x\big) }{\sqrt{2}}.
\end{equation}
As is shown in Appendix \ref{B} the solution of (\ref{eq:25}) is as follows:

\begin{eqnarray}
    \label{eq:q146}
    \begin{cases}
    G_{11}^{R}(u,v;u^{\prime},v^{\prime})=-\dfrac{\Phi(v^{\prime})}{2}J_{0}\bigg(2\sqrt{(u-u^{\prime})a(v,v^{\prime})}\bigg)\theta_{uu^{\prime}}\theta_{vv^{\prime}} \ , \\
    G_{22}^{R}(u,v;u^{\prime},v^{\prime})=-\dfrac{\Phi(v)}{2}J_{0}\bigg(2\sqrt{(u-u^{\prime})a(v,v^{\prime})}\bigg)\theta_{uu^{\prime}}\theta_{vv^{\prime}} \ , \\
    G_{12}^{R}(u,v;u^{\prime},v^{\prime})=-\dfrac{i}{\sqrt{2}}\delta_{uu^{\prime}}\theta_{vv^{\prime}} + \dfrac{i}{\sqrt{2}}\sqrt{\dfrac{a(v,v^{\prime})}{u-u^{\prime}}}J_{1}\bigg(2\sqrt{(u-u^{\prime})a(v,v^{\prime})}\bigg)\theta_{uu^{\prime}}\theta_{vv^{\prime}} \ , \\
    G_{21}^{R}(u,v;u^{\prime},v^{\prime})=-\dfrac{i}{\sqrt{2}}\theta_{uu^{\prime}}\delta_{vv^{\prime}} + \dfrac{i}{\sqrt{2}}\dfrac{\Phi(v)\Phi(v^{\prime})}{2}\sqrt{\dfrac{u-u^{\prime}}{a(v,v^{\prime})}}J_{1}\bigg(2\sqrt{(u-u^{\prime})a(v,v^{\prime})}\bigg)\theta_{uu^{\prime}}\theta_{vv^{\prime}}.
    \end{cases}
\end{eqnarray}
Here for simplicity we denote $\theta_{xy} \equiv \theta(x-y)$ and $\delta_{xy} \equiv \delta(x-y)$,

\begin{equation}\label{avv}
    a(v,v') = \frac12 \, \int_{v'}^v dy \, \Phi^2(y);
\end{equation}
and $J_0(x)$, $J_1(x)$ are the Bessel functions.

It is straightforward to check that when $\Phi(v)=0$, the obtained expression reduces to:
\begin{equation}
    \label{eq:45}
    G^{R}\bigg|_{\Phi=0}=\left[ {\begin{array}{cc}
   0 & -\dfrac{i}{\sqrt{2}}\theta(v-v^{\prime})\delta(u-u^{\prime}) \\
    -\dfrac{i}{\sqrt{2}}\delta(v-v^{\prime})\theta(u-u^{\prime}) & 0 \\
  \end{array} } \right]=G^{R}_{0} \ ,
\end{equation}
which is the retarded Green function in the absence of the background scalar field, as is explained in Appendix \ref{A}.
The advanced Green function can be obtained via the Hermitian conjugation of the retarded one:

\begin{equation}
    \label{eq:46}
    G^{A}={\big[G^{R}\big]}^{\dagger},
\end{equation}
Hermitian conjugation includes complex conjugation along with the interchange of the arguments $u,v \leftrightarrow u^{\prime},v^{\prime}$.

The knowledge of the exact form of $G^{R/A}$ allows one to find the Keldysh propagator from \eqref{eq:51}. This form is bulky and hard to treat in physical terms. That is the reason why we are interested in the calculation of the expectation values of the scalar current, $\langle \bar{\psi} \psi \rangle$, and of the stress energy tensor. Then, one only needs the trace of the Keldysh propagator over the spinor indexes for coincident points.

\subsection{The expectation value of the scalar current}

The physical observables that we calculate below can be expressed in terms of the Wightman propagator $G^{<}(x,t;x',t') = i \braket{\bar{\psi}(t,x)\psi(t',x')}$, which can be represented as

\begin{equation}
    \label{eq:o1}
   G^{<}(x,t;x^{\prime},t^{\prime})=\dfrac{1}{2}G^{K}(x,t;x^{\prime},t^{\prime})-\dfrac{1}{2}\bigg(G^{R}(x,t;x^{\prime},t^{\prime})-G^{A}(x,t;x^{\prime},t^{\prime})\bigg) \ .
\end{equation}
As we will see, the second term on the right hand side of this expression does not contribute to the quantities that we calculate, at least in the limit that we describe below in this section.

Thus, for the beggining we need to find the exact $G^{K}(t,x;t^{\prime},x^{\prime})$ propagator when $t^{\prime}=t$ and $x^{\prime}=x-\delta$, as $\delta \to 0$. We split $x$ and $x^{\prime}$ to single out the divergent part of the correlation function. As we will see below, the divergent part is only in the diagonal components of $G^{K}_{th}$.

We start with the calculation of $G^{K}_{th}(x,t;x,t)$:

\begin{eqnarray}
    \label{eq:55}
     \lim_{\delta \to 0}G^{K}_{th}(x,t;x-\delta,t) = \nonumber \\
     =-\lim_{\delta \to 0 } \sqrt{2}\int_{-\infty}^{+\infty} d\tau f(\sqrt{2}\tau)\bigg(G^{R}(x,t;x-\delta,t-\sqrt{2}\tau)+G^{A}(x,t-\sqrt{2}\tau;x-\delta,t)\bigg) \ ,
\end{eqnarray}
where we have made a change of integration variables in eq.(\ref{eq:51}) and $f\big(\tau\big)$ is defined by eq. (\ref{eq:52}).

Let us start with the analyzes of the structure of $[G^{K}_{th}]_{11}$, which is the $11$ component of the spinor matrix $G_{th}^K$. From \eqref{eq:q146} and \eqref{eq:46} we find that:

\begin{equation}
    \label{eq:56}
    G_{11}^{R}(x,t;x-\delta,t-\sqrt{2}\tau)+G_{11}^{A}(x,t-\sqrt{2}\tau;x-\delta,t)=
\end{equation}
$$=-\dfrac{1}{2}\theta(\tau-\delta)J_{0}\bigg(2\sqrt{\tau a(v,v-\tau)}\bigg)\bigg[\Phi\big(v-\tau\big)+\Phi\big(v\big)\bigg].$$
Using the properties of the retarded and advanced propagators we find that:

\begin{eqnarray}
    \label{eq:61}
    [G^{K}_{th}]_{11}(x,t;x,t) = [G^{K}_{th}]_{22}(x,t;x,t) \ , \nonumber \\
    G^{R}_{12}(x,t;x,\tau)+G^{A}_{12}(x,\tau;x,t)=0 \ , \nonumber \\
     G^{R}_{21}(x,t;x,\tau)+G^{A}_{21}(x,\tau;x,t)=0 \ .
\end{eqnarray}
As a result the spinor matrix $G^{K}_{th}(x,t;x,t)$ has the following form:

\begin{equation}
    \label{eq:62}
     G^{K}_{th}(x,t;x-\delta,t)=\sqrt{2}\int_{\delta}^{+\infty} d\tau f\big(\sqrt{2}\tau\big)J_{0}\bigg(2\sqrt{\tau a(v,v-\tau)}\bigg)\dfrac{1}{2}\bigg[\Phi\big(v-\tau\big)+\Phi\big(v\big)\bigg]
    \left[ {\begin{array}{cc}
   1 & 0 \\
    0 & 1 \\
  \end{array} } \right].
\end{equation}
Since the Bessel function obeys the following property $J_{0}(x) \approx 1$, as $x \rightarrow{} 0$, the propagator $G^{K}_{th}$ has the logarithmic divergence $\log(\delta)$ at the coincidence limit, $\delta \to 0$. We come back to this point below.

Now we continue with the discussion of $G^K_{an}$ in \eqref{eq:51}. In this case one can put $\delta=0$, since there is no divergence in the coincidence limit in such a term. The calculation of $G^K_{an}$ is given in the Appendix \ref{C}. The result of the calculation is:

\begin{equation}\label{GKth}
    G^K_{an}(x,t;x,t) = \bigg\{[G^K_{an2}]_{11}(x,t;x,t) + [G^K_{an3}]_{11}(x,t;x,t)\bigg\} \, \left[ {\begin{array}{cc}
   1 & 0 \\
    0 & 1 \\
  \end{array} } \right],
\end{equation}
where $[G^K_{an2}]_{11}(x,t;x,t)$ is defined in the eq. (\ref{eq:163}), while $[G^K_{an2}]_{11}(x,t;x,t)$ is defined in eq. (\ref{eq:166}).

In all, we express the expectation value of the scalar current as:

\begin{eqnarray}
    \label{eq:SETF2}
2i\lim_{t\to t', x\to x'}\braket{\bar{\psi}(t^{\prime},x^{\prime})\,\psi(t,x)} = Tr \lim_{t\to t', x\to x'}\left\{ G^{K}_{th}(x,t;x^{\prime},t^{\prime}) +  G^{K}_{an2}(x,t;x^{\prime},t^{\prime}) \right. + \nonumber \\ \left. + G^{K}_{an3}(x,t;x^{\prime},t^{\prime}) - \left[G^{R}(x,t;x^{\prime},t^{\prime})-G^{A}(x,t;x^{\prime},t^{\prime})\right]\right\},
\end{eqnarray}
where the spinor matrixes $G^K_{an2}$ and $G^K_{an3}$ are defined in eq. (\ref{eq:d168}) and eq. (\ref{eq:e170}) in Appendix \ref{C}, correspondingly.

In principle using the exact form of $G^K$ at coincident points one can express the current via integrals over the Bessel functions. But to simplify expressions and to reduce them to a physically tractable form, we make the following assumptions. First, we put $T=0$, then the distribution function becomes:

\begin{equation}
    \label{eq:k1}
   f(\tau)=- \lim_{T\to 0}\mathcal{P}\dfrac{iT}{\sinh{\big(\pi T \tau\big)}}=-\mathcal{P}\dfrac{i}{\pi \tau} \ .
\end{equation}
Second, we look for $\langle \bar{\psi} \psi \rangle$ for such values of $\Phi(v)$, which obey the following conditions:

\begin{equation}
    \label{eq:100}
    \dfrac{\Phi(v)}{\lambda} \gg 1, \qquad \qquad \bigg|\dfrac{\Phi^{(k)}(v)}{\lambda^{k} \Phi(v)}\bigg| \ll 1, \quad {\rm and} \quad \dfrac{1}{\lambda}\int_{v-1}^{v}\Phi^{2}(y)dy \gg 1,
\end{equation}
which essentially means that the scalar field is large and slowly changing function. Our goal is to find the leading contribution to the $\braket{\bar{\psi}(t,x-\delta)\psi(t,x)}$ in the limit (\ref{eq:100}).

The details of the calculation one can find in the Appendix \ref{D}. The result is

\begin{equation}
    \label{eq:13000}
    \langle\bar{\psi}(t,x-\delta)\psi(t,x)\rangle \approx \dfrac{\Phi(v)}{\pi}\ln\bigg[\Phi(v) \, \delta \bigg].
\end{equation}
We discuss the meaning of this result in the concluding section.

\subsection{The expectation value of the stress energy tensor}

The stress energy operator in the theory under consideration is as follows:

\begin{equation}
    \label{eq:SET8}
    T^{\mu \nu}=\dfrac{i}{4}\bigg[\bar{\psi}\gamma^{\mu}\partial^{\nu}\psi+\bar{\psi}\gamma^{\nu}\partial^{\mu}\psi-\partial^{\nu}\bar{\psi}\gamma^{\mu}\psi-\partial^{\mu}\bar{\psi}\gamma^{\nu}\psi\bigg]-\eta^{\mu \nu} \mathcal{L} \ .
\end{equation}
The component of this tensor, which is the most interesting for us is $T^{01}$ --- the flux. We will concentrate on the calculation of this component in the present subsection. The diagonal components of $T^{\mu\nu}$ will be discussed in the next section.

After the point splitting regularisation we can represent the expectation value of the flux via the derivatives of the components of the propagator \eqref{eq:SETF2}:

\begin{equation}
    \label{eq:SETF1}
    \braket{T^{01}(t,x)}=\dfrac{i}{4}\bigg[\big(\sqrt{2}\partial_{v}-\sqrt{2}\partial_{v^{\prime}}\big)\braket{\bar{\psi}_{1}(t^{\prime},x^{\prime})\psi_{2}(t,x)}-\big(\sqrt{2}\partial_{u}-\sqrt{2}\partial_{u^{\prime}}\big)\braket{\bar{\psi}_{2}(t^{\prime},x^{\prime})\psi_{1}(t,x)}\bigg]\bigg|_{t,x=t^{\prime},x^{\prime}} \ .
\end{equation}
We will denote the contribution of each term from \eqref{eq:SETF2} into $\braket{T^{01}(t,x)}$ as $T_{G^{K}_{th}},T_{G^{K}_{an1}}$ and etc. \ .

Since calculations are straightforward, but tedious, we show here only the final results:

\begin{eqnarray}
\label{eq:SETF7}
    T_{G^{R}-G^{A}}=0, \quad
    T_{G^K_{an2}}=-T_{G^K_{th}}-T_{G^{K}_{an3}},  \quad
    T_{G^{K}_{an3}} = - T_{G^{K}_{th}}- \nonumber \\
    -\dfrac{1}{2}\int dy \int d\tau_{1} d\tau_{2} f\big(\tau_{1}-\tau_{2}\big) \bigg[\Phi(v_{\tau_{2}})-\Phi(v_{\tau_{1}})\bigg]\theta(v-v_{\tau_{1}})\theta(u-u_{\tau_{1}})J_{1}\bigg(2\sqrt{a(v,v_{\tau_{1}})(u-u_{\tau_{1}})}\bigg)\times \nonumber \\
    \times \bigg[\dfrac{\Phi(v_{\tau_{1}})\Phi^{2}(v)}{4}\sqrt{\dfrac{u-u_{\tau_{1}}}{a(v,v_{\tau_{1}})}}\theta(v-v_{\tau_{2}})\delta(u-u_{\tau_{2}})-\dfrac{\Phi(v)}{2}\sqrt{\dfrac{a(v,v_{\tau_{1}})}{u-u_{\tau_{1}}}}\delta(v-v_{\tau_{2}})\theta(u-u_{\tau_{2}})\bigg],
    \nonumber \\
    T_{G^{K}_{th}}=\dfrac{1}{2\sqrt{2}}\int_{0}^{\infty} d\tau f\big(\sqrt{2}\tau\big)\bigg(\Phi(v)-\Phi(v-\tau)\bigg)^{2}J_{0}\bigg(2\sqrt{\tau a(v,v-\tau)}\bigg).
\end{eqnarray}
To obtain these expressions we have used the property explained around eq. \eqref{eq: ,161} from the Appendix \ref{C} and the following property of the Bessel function \cite{9}: \  $\dfrac{2n}{x}J_{n}(x)=J_{n-1}(x)+J_{n+1}(x)$.
Putting all equations \eqref{eq:SETF7} together, we get that the non diagonal component of the stress energy tensor is zero:
\begin{equation}
    \label{eq:SETF9}
    \braket{T^{01}(t,x)}=0,
\end{equation}
which means that there is no fermion flux in the background field under consideration for the given state in question. This result follows from the fact that in the functional formalism and the corresponding state the left and right moving fermions always enter symmetrically. We will see below that for some other states the situation can be different.

\section{Operator formalism and the corresponding simplest state}\label{45}

In the previous section we have calculated propagators and expectation values of physical quantities using solutions of the Dyson-Schwinger equation corresponding to an initial (thermal or ground) state at past infinity, which is the simplest in those settings. Now we want to calculate the same quantities in the operator formalism also using an initial state, which is the simplest in new settings. Along this way we will encounter subtleties in matching the results for the physical observables in the two formalisms under discussion. We will argue that the discrepancy comes from the fact that in different formalisms we loosely consider different states, which have physically distinct properties. Then, in the section 4 we propose a way to match the states in the two cases.

\subsection{Modes and canonical commutation relation}

To quantize the theory in operator formalism, we start from the equations of motion for the modes following from the action (\ref{eq:2}):

\begin{equation}
    \label{eq:131}
  \left[ {\begin{array}{cc}
   -\Phi(v) & i\sqrt{2}\partial_{u} \\
    i\sqrt{2}\partial_{v} & -\Phi(v) \\
  \end{array} } \right]\begin{bmatrix}\psi_{1} \\ \psi_{2} \end{bmatrix}=0 \ .
\end{equation}
Two linear independent solutions of this equation are as follows:

\begin{equation}
    \label{eq:132}
    \chi_p(t,x) = \tilde{u}(p)e^{-ipu-ia(v,0)/p} \qquad \text{and} \qquad  \zeta_p(t,x) = \tilde{v}(p)e^{ipu+ia(v,0)/p}  \ ,
\end{equation}
where $v$ and $u$ we define in eq.(\ref{LCcoord}),  $a(v,v^{\prime})$ is defined in eq. \eqref{avv} and

\begin{equation}
    \label{eq:133}
     \tilde{u}(p)=\begin{bmatrix}1 \\ \dfrac{\Phi(v)}{\sqrt{2}p} \end{bmatrix} \qquad \text{and} \qquad  \tilde{v}(p)=\begin{bmatrix}1 \\ -\dfrac{\Phi(v)}{\sqrt{2}p} \end{bmatrix}  \ .
\end{equation}
We choose such modes, as in \eqref{eq:132}, \eqref{eq:133}, because in the case when $\Phi(v)=m$ the correspodning Keldysh propagator is the same as for the standard plane waves, as we show in Appendix \ref{E}. In fact, we show there that if we write the field operator\footnote{The form of the field operator under consideration is formal, because of the peculiarity of the modes at $p=0$. That is why in the calculations of observables we assume that there is an $\epsilon$ shift in the exponents: $u \rightarrow u - i \epsilon, \quad v \rightarrow v -i\epsilon.$} as:

\begin{equation}
    \label{eq:134}
    \hat{\psi}(u,v)=\int_{0}^{+\infty}\dfrac{dq}{2\pi}\dfrac{1}{\sqrt[4]{2}}\bigg[\hat{a}_{q}\begin{bmatrix}1 \\ \dfrac{\Phi(v)}{\sqrt{2}q} \end{bmatrix}e^{-iqu-ia(v,0)/q}+\hat{b}_{p}^{\dagger}\begin{bmatrix}1 \\ -\dfrac{\Phi(v)}{\sqrt{2}q} \end{bmatrix}e^{iqu+ia(v,0)/q}\bigg] \ ,
\end{equation}
with

\begin{equation}
    \label{eq:135}
 \{\hat{a}_{p},\hat{a}_{q}^{\dagger}\}=2\pi \delta(p-q), \quad
\{\hat{b}_{p},\hat{b}_{q}^{\dagger}\}=2\pi \delta(p-q),
\end{equation}
and then put $\Phi(v)=m$ the standard theory of massive fermions follows after a trivial Bogolyubov transformation, which does not mix positive and negative modes.

The state that we are going to use in our calculations below is the ground Fock space state $\hat{a}_p |0\rangle = \hat{b}_p |0\rangle = 0$. We come back to the discussion of other possible states in the next section.

\subsection{Retarded and advanced Green functions}

As we have explained above, the retarded and advanced Green functions in Gaussian approximation do not dependent on the choice of initial state. In fact, the retarded Green function is equal to:

\begin{equation}
    \label{eq:138}
    G^{R}(u,v;u^{\prime},v^{\prime}) \equiv -i \theta(t-t^{\prime})\{\psi(u,v),\bar{\psi}(u^{\prime},v^{\prime})\},
\end{equation}
where the anti--commutator of $\psi$'s is the c--number. The independence of the state is apparent for the tree--level propagators for the case when $\Phi=0$. In this subsection we observe the same fact for the exact retarded and advanced propagators in the Gaussian approximation (see e.g. \cite{Kamenev} and \cite{Akhmedov:2013vka}).

From (\ref{eq:138}) and (\ref{eq:134}) for $t>t^{\prime}$ we obtain that:

\small
\begin{eqnarray}
    \label{eq:140}
    iG^{R}(u,v;u^{\prime},v^{\prime})=\nonumber \\
    \left[ {\begin{array}{cc}
   \dfrac{-i\Phi(v^{\prime})}{2}\int_{0}^{+\infty}\dfrac{dp}{\pi}\dfrac{1}{p}\sin{\bigg(p(u-u^{\prime})+a(v,v^{\prime})/p\bigg)} & \dfrac{1}{\sqrt{2}}\int_{0}^{+\infty}\dfrac{dp}{\pi}\cos{\bigg(p(u-u^{\prime})+a(v,v^{\prime})/p\bigg)} \\
    \dfrac{\Phi(v)\Phi(v^{\prime})}{2\sqrt{2}}\int_{0}^{+\infty}\dfrac{dp}{\pi}\dfrac{1}{p^{2}}\cos{\bigg(p(u-u^{\prime})+a(v,v^{\prime})/p\bigg)} & \dfrac{-i\Phi(v)}{2}\int_{0}^{+\infty}\dfrac{dp}{\pi}\dfrac{1}{p}\sin{\bigg(p(u-u^{\prime})+a(v,v^{\prime})/p\bigg)} \\
  \end{array} } \right].
\end{eqnarray}
\normalsize
The integrals here are equal to \cite{1}, \cite{12}:

\begin{eqnarray}
    \label{eq:141}
    -\theta(t-t^{\prime})\dfrac{i\Phi(v)}{2}\int_{0}^{+\infty}\dfrac{dp}{\pi}\dfrac{1}{p}\sin{\bigg(p(u-u^{\prime})+a(v,v^{\prime})/p\bigg)}=\nonumber \\
    =\dfrac{-i\Phi(v)}{2}J_{0}\bigg(2\sqrt{(u-u^{\prime})a(v,v^{\prime})}\bigg)\theta(u-u^{\prime})\theta(v-v^{\prime}),
\end{eqnarray}
and:

\begin{eqnarray}
    \label{eq:142}
    \theta(t-t^{\prime})\int_{0}^{+\infty}\dfrac{dp}{\pi}\cos{\bigg(p(u-u^{\prime})+a(v,v^{\prime})/p\bigg)}=\nonumber \\
    =-\sqrt{\dfrac{a(v,v^{\prime})}{u-u^{\prime}}}J_{1}\bigg(2\sqrt{(u-u^{\prime})a(v,v^{\prime})}\bigg)\theta(u-u^{\prime})\theta(v-v^{\prime}), \qquad \text{where} \qquad  u-u^{\prime} \neq 0 \ .
\end{eqnarray}
At the same time, when $u-u^{\prime}=0$ one has that:

\begin{equation}
    \label{eq:143}
    \theta(t-t^{\prime})\int_{0}^{+\infty}\dfrac{dp}{\pi}\cos{\bigg(a(v,v^{\prime})/p\bigg)}=\theta(t-t^{\prime})\int_{0}^{+\infty}\dfrac{dp}{\pi}\dfrac{1}{p^{2}}\cos{\bigg(pa(v,v^{\prime})\bigg)}=
\end{equation}
$$=\theta(v-v^{\prime})\lim_{\epsilon \to 0}\dfrac{1}{\pi\epsilon}-\dfrac{1}{2}a(v,v^{\prime})\theta(v-v^{\prime}) \ .$$
Then it is possible to express the result of the integration as:

\begin{equation}
    \label{eq:144}
    \theta(t-t^{\prime})\int_{0}^{+\infty}\dfrac{dp}{\pi}\cos{\bigg(p(u-u^{\prime})+a(v,v^{\prime})/p\bigg)}=
\end{equation}
$$=\theta(v-v^{\prime}) \delta(u-u^{\prime})-\sqrt{\dfrac{a(v,v^{\prime})}{u-u^{\prime}}}J_{1}\bigg(2\sqrt{(u-u^{\prime})a(v,v^{\prime})}\bigg)\theta(u-u^{\prime})\theta(v-v^{\prime}) \ .$$
Integral in the $G^{R}_{21}$ component of the spinor matrix $G^R$ can be reduced to the same form as \eqref{eq:142}:

\begin{equation}
    \label{eq:145}
    \dfrac{\Phi(v)\Phi(v^{\prime})}{2}\int_{0}^{+\infty}\dfrac{dp}{\pi}\dfrac{1}{p^{2}}\cos{\bigg(p(u-u^{\prime})+a(v,v^{\prime})/p\bigg)}=
\end{equation}
$$=\theta(u-u^{\prime}) \delta(v-v^{\prime})-\dfrac{\Phi(v)\Phi(v^{\prime})}{2}\sqrt{\dfrac{u-u^{\prime}}{a(v,v^{\prime})}}J_{1}\bigg(2\sqrt{(u-u^{\prime})a(v,v^{\prime})}\bigg)\theta(u-u^{\prime})\theta(v-v^{\prime}).$$
Finally, putting all these results together we find that the retarded Green function has the same form as (\ref{eq:q146}).

\subsection{The expectation value of the scalar current}

The Keldysh propagator is equal to:

 \begin{equation}
     \label{ft21}
     G^{K} (t^{\prime},x^{\prime}, t,x)= - i \langle 0| \big[\hat{\psi}(t^{\prime},x^{\prime}), \hat{\bar{\psi}}(t,x)\big] |0\rangle=-i \int_{0}^{\infty}\dfrac{dp}{2\pi}\bigg(\chi_{p}(t^{\prime},x^{\prime})\bar{\chi}_{p}(t,x) - \zeta_{p}(t^{\prime},x^{\prime}) \bar{\zeta}_{p}(t,x)\bigg),
 \end{equation}
where the spinor modes $\chi_p$ and $\zeta_p$ are defined in (\ref{eq:132}) and (\ref{eq:133}). In the expectation value under consideration we use the state $|0\rangle$, which is annihilated by $\hat{a}_p$ and $\hat{b}_p$ operators from eq. (\ref{eq:134}).

Using that:

\begin{eqnarray}
    \label{eq:152}
    \int_{0}^{+\infty}\dfrac{dp}{2\pi}\dfrac{1}{p}\cos{\bigg(p(u-u^{\prime})+a(v,v^{\prime})/p\bigg)}=\nonumber \\
    =\begin{cases}
    -\dfrac{1}{2}N_{0}\bigg(2\sqrt{(u-u^{\prime})a(v,v^{\prime})}\bigg), \qquad \ \  (u-u^{\prime})(v-v^{\prime})>0\\
    \ \ \dfrac{1}{\pi}K_{0}\bigg(2\sqrt{|(u-u^{\prime})a(v,v^{\prime})|}\bigg), \qquad (u-u^{\prime})(v-v^{\prime})<0 \ ,
    \end{cases}
\end{eqnarray}
we can express the trace of the spinor matrix $G^K$, which we need to calclate the scalar current, as:

 \begin{eqnarray}
     \label{eq:153}
      Tr\bigg( G^{K}(u,v;u^{\prime},v^{\prime}) \bigg)=-i\bigg(\Phi(v)+\Phi(v^{\prime})\bigg) \int_{0}^{+\infty}\dfrac{dp}{2\pi}\dfrac{1}{p}\cos{\bigg(p(u-u^{\prime})+a(v,v^{\prime})/p\bigg)}=\nonumber \\
      =-i\bigg(\Phi(v)+\Phi(v^{\prime})\bigg)\bigg[-\theta\big((u-u^{\prime})(v-v^{\prime})\big)\dfrac{1}{2}N_{0}\bigg(2\sqrt{(u-u^{\prime})a(v,v^{\prime})}\bigg)+\nonumber \\
      +\theta\big(-(u-u^{\prime})(v-v^{\prime})\big)\dfrac{1}{\pi}K_{0}\bigg(2\sqrt{|(u-u^{\prime})a(v,v^{\prime})|}\bigg)\bigg],
 \end{eqnarray}
where $K_0(x)$ is the MacDonald function and $N_0(x)$ is the Neumann function.

From the obtained expression one can find that in the limit (\ref{eq:100}) the scalar current is equal to:

\begin{eqnarray}
    \label{eq:154}
    \lim_{\delta\to 0}\langle 0| \bar{\psi}(t+\delta^{0},x+\delta^{1})\psi(t,x)|0\rangle = \frac{-i}{2} \, \lim_{\delta\to 0} Tr\bigg[G^{K}\bigg(u,v;u+\delta^{u},v+\delta^{v}\bigg)\bigg] = \nonumber \\
    = -\frac12 \lim_{\delta\to 0} \bigg(\Phi(v)+\Phi(v+\delta^{v})\bigg)\bigg[-\theta\big(\delta^{v}\delta^{u}\big)\dfrac{1}{2}N_{0}\bigg(\Phi(v)\delta\bigg)+\theta\big(-\delta^{v}\delta^{u}\big)\dfrac{1}{\pi}K_{0}\bigg(\Phi(v)\delta\bigg)\bigg] \approx \nonumber \\
    \approx \dfrac{\Phi(v)}{\pi} \,\ln\bigg[\Phi(v) \, \delta\bigg],
\end{eqnarray}
where $\delta \equiv \sqrt{\delta_{\mu}\delta^{\mu}}=\sqrt{2\delta^{v}\delta^{u}}$. This expression coincides with (\ref{eq:13000}) --- with the scalar current calculated in the functional formalism in the same limit. We explain these observations in the concluding section.

\subsection{The expectation value of the stress energy tensor}

In this subsection we discuss the expectation value of the stress energy operator (\ref{eq:SET8}) in the operator formalism. Since the modes in \eqref{eq:134} satisfy equations of motion, we can skip $\mathcal{L}$ in the formula above. In the Appendix \ref{F} we perform the calculation of the expectation value using the expression (\ref{ft21}) for the Keldysh propagator. The result of the calculation in the limit (\ref{eq:100}) is as follows:

\begin{equation}
    \label{e:35}
   \langle T^{\mu \nu} \rangle_{reg} \approx \dfrac{\Phi^{2}(v)}{2\pi}\, \ln{\bigg[\Phi(v) \, \delta\bigg]} \, \eta^{\mu \nu}-\dfrac{1}{48\pi}\{\bar{a}(v),v\}
   \left[ \begin{array}{cc}
   1& 1 \\
    1 & 1 \\
  \end{array}  \right] \ ,
\end{equation}
where

$$
\bar{a}(v)=\dfrac{1}{m^{2}}\int_{}^{v}dy\Phi^{2}(y),
$$
and $\{f(z),z\}$ is the Schwarzian derivative:
\begin{equation}
        \{f(z),z\}=\dfrac{f^{\prime \prime \prime}(z)}{f^{\prime}(z)}-\dfrac{3}{2}\bigg(\dfrac{f^{\prime \prime}(z)}{f^{\prime}(z)}\bigg)^{2} \nonumber .
\end{equation}
The second term in \eqref{e:35} breaks the general covariance. But that is not surprising since we consider the theory with the position dependent potential $\Phi(t-x)$. Note that for the case $\Phi=m$ the general covariance in (\ref{e:35}) is restored.

The presence of the first term in \eqref{e:35} can be explained as follows. From the definition of the stress energy operator (\ref{eq:SET8}), with the help of the equations of motion \eqref{eq:131}, one can deduce that

\begin{equation}
    \label{ey1}
    \langle 0| T^{\mu}_{\mu} |0 \rangle=\Phi(v)\langle 0| \bar{\psi}(t,x)\psi(t,x)|0\rangle.
\end{equation}
Hence, the expectation value of the stress energy tensor should contain the term as follows:

\begin{equation}
    \label{ey2}
    \langle 0| T^{\mu \nu} |0 \rangle=\dfrac{1}{2}\Phi(v)\langle 0| \bar{\psi}(t,x)\psi(t,x)|0\rangle \eta^{\mu \nu}+...=\dfrac{\Phi^{2}(v)}{2\pi}\, \ln{\bigg[\Phi(v)\,\delta\bigg]} \, \eta^{\mu \nu}+... \ ,
\end{equation}
which is a part of the diagonal component of the expectation value. The same term is present in the diagonal components of the stress energy tensor, which was found in the functional formalism. As we explain in the concluding section, this term is universal in the limit (\ref{eq:100}).

A bit more unusual is the presence of the second Schwarzian term in (\ref{e:35}). In particular the existence of this term means that there is non--zero fermion flux $\langle T^{01}\rangle$, which was absent in the functional formalism. So we encounter here a physically distinct state. The situation is somewhat similar to the one with Boulewar and Hartle--Hawking vs. Unruh states in the presence of eternal black holes \cite{Candelas:1980zt}.

To understand the origin of the Schwarzian term in \eqref{e:35}, consider the action of our theory

\begin{equation}
    \label{eq:cft4}
    S=\int du dv \bigg[\psi^{\dagger}_{1}i \sqrt{2} \partial_{v}\psi_{1} + \psi^{\dagger}_{2}i \sqrt{2} \partial_{u}\psi_{2}-\Phi(v)\big(\psi^{\dagger}_{1}\psi_{2}+\psi^{\dagger}_{2}\psi_{1}\big) \bigg] \ ,
\end{equation}
and perform the following transformation in it:

\begin{equation}
    \label{eq:cft5}
    \begin{cases}
    v^{\prime}=\dfrac{1}{m^{2}}\int_{0}^{v}dy \ \Phi^{2}(y) \\
    u^{\prime}=u
    \end{cases} \qquad \text{and} \qquad
     \begin{cases}
    \psi_{1}(u,v)=\psi_{1}(u^{\prime},v(v^{\prime}))=\eta_{1}(u^{\prime},v^{\prime}) \\
    \psi_{2}(u,v)=\sqrt{\dfrac{dv^{\prime}}{dv}}\psi_{2}(u^{\prime},v(v^{\prime}))=\sqrt{\dfrac{dv^{\prime}}{dv}}\eta_{2}(u^{\prime},v^{\prime}) \ .
    \end{cases}
\end{equation}
This is the conformal map, but we perform it in the theory which is not conformally invariant, due to the presence of the interaction term.

Under such a transformation the action (\ref{eq:cft4}) gets converted into

\begin{eqnarray}
\label{eq:cft7}
S = \int du^{\prime} dv^{\prime} \bigg[\eta^{\dagger}_{1}i\sqrt{2}\partial_{v^{\prime}}\eta_{1}+\eta^{\dagger}_{2}i\sqrt{2}\partial_{u^{\prime}}\eta_{2}-m\big(\eta^{\dagger}_{1}\eta_{2}+\eta^{\dagger}_{2}\eta_{1}\big)\bigg] \ ,
\end{eqnarray}
which is the theory of free massive fermions in Minkowski space. Furthermore, under such a transformation the field operator $\hat{\psi}(u,v)$, which initially is defined as (\ref{eq:134}),
gets converted into:

\begin{eqnarray}
    \label{eq:cft9}
    \hat{\eta}(u^{\prime},v^{\prime})=\begin{bmatrix} \eta_{1} \\ \eta_{2} \end{bmatrix}=\int_{0}^{+\infty}\dfrac{dq}{2\pi}\dfrac{1}{\sqrt[4]{2}}\bigg[\hat{a}_{q}\begin{bmatrix}1 \\ \dfrac{m}{\sqrt{2}q} \end{bmatrix}e^{-iqu^{\prime}-im^{2}v^{\prime}/2q}+\hat{b}_{p}^{\dagger}\begin{bmatrix}1 \\ -\dfrac{m}{\sqrt{2}q} \end{bmatrix}e^{iqu^{\prime}+im^{2}v^{\prime}/2q}\bigg]= \nonumber \\
    =\int_{-\infty}^{+\infty}\dfrac{dp}{2\pi}\bigg[\hat{\tilde{a}}_{p}u(p)e^{-iw_{p}t^{\prime}+ipx^{\prime}}+\hat{\tilde{b}}_{p}^{\dagger}v(p)e^{iw_{p}t^{\prime}-ipx^{\prime}}\bigg]
\end{eqnarray}
where we use the results and notations of Appendix \ref{E}.

Thus, under the conformal transformation in question the theory with the potential $\Phi$ gets converted into the theory of free massive fermions with the standard plane wave modes. At the same time, we show in Appendix \ref{G} that under such transformations as (\ref{eq:cft5}) the expectation value of the flux operator $\langle T^{01}\rangle$ gets shifted affinely by the Schwarzian derivative type of contribution. Meanwhile, the theory (\ref{eq:cft7}) obviously has zero flux\footnote{After all the appearance of the Schwarzian term is not so surprising, because the flux operator $T^{01}$ in 2D generates the Virasoro algebra independently of whether the theory is conformaly invariant or not. We would like to thank G.Jeorgadze for pointing to us out this fact.}. These observations explain the origin of the Schwarzian contribution to the expectation value of the stress energy tensor.

\section{The relation between previously found Kledysh propagators}

Thus, in the section 2 we have solved the Dyson-Schwinger equation and found the exact retarded and advanced propagators in the functional formalism. Then we have calculated the physical observables following from the exact Keldysh propagator \eqref{eq:51}, which carries the information about the state of the theory.

In the section \ref{45}, we have been working in the operator formalism and found a complete basis of modes solving the classical equations of motion. But there is an ambiguity in the choice of such a basis. Depending on this choice, there are different ``ground'' Fock space states in the theory. In fact, instead of (\ref{eq:132}) and (\ref{eq:133}) one could consider canonically transformed basis of modes:

\begin{equation}
    \label{bt02}
    \widetilde{\chi}_{p}(t,x) = \int \dfrac{dq}{2\pi}\bigg[\alpha_{pq}\chi_{q}(t,x) + \beta_{pq} \zeta_{q}(t,x)\bigg], \quad
    \widetilde{\zeta}_{p}(t,x) = \int \dfrac{dq}{2\pi} \bigg[\gamma_{pq}\chi_{q}(t,x) + \eta_{pq}\zeta_{q}(t,x)\bigg].
\end{equation}
To respect the canonical anti--commutation relations for the fermionic fields and for the corresponding creation and annihilation operators the Bogoliubov coefficients, $\alpha_{pq}, \beta_{pq}, \gamma_{pq}$ and $\eta_{pq}$, should satisfy the following conditions:

\begin{eqnarray}
    \int_{}^{}\dfrac{dp}{2\pi}\bigg(\alpha_{pq} \alpha^{*}_{pq^{\prime}} + \gamma_{pq} \gamma^{*}_{pq^{\prime}}\bigg) = 2\pi \delta(q-q^{\prime}), \quad
    \int_{}^{}\dfrac{dp}{2\pi}\bigg(\beta^{*}_{pq} \beta_{pq^{\prime}} + \eta^{*}_{pq} \eta_{pq^{\prime}}\bigg)=2\pi \delta(q-q^{\prime}), \nonumber \\
    {\rm and} \quad \int_{}^{}\dfrac{dp}{2\pi}\bigg(\alpha_{pq} \beta^{*}_{pq^{\prime}} + \gamma_{pq} \eta^{*}_{pq^{\prime}}\bigg)=0.
\end{eqnarray}
On physical grounds one also should demand that

\begin{eqnarray} \label{alphapq}
    \alpha_{pq} \approx \eta_{pq} \approx \delta(p-q), \quad \beta_{pq} \approx \gamma_{pq} \approx 0,
\end{eqnarray}
as either $p$ or $q$ is taken to infinity. That is necessary for the modes to have the proper UV behavior and correspondingly for the propagators to have the proper Hadamard behaviour or UV singularity structure.

Thus, there is no unique way to choose the basis of modes and all possibilities in (\ref{bt02}) are in principle allowed and may lead to different physical situations. This fact is apparent when there is no preferable basis of special functions, which is obviously the case in the present situation.

For a given choice of modes one can define a new Fock space ``ground'' state:

\begin{equation}
    \hat{\widetilde{a}}_p |\alpha, \beta,\gamma,\eta\rangle = \hat{\widetilde{b}}_p |\alpha, \beta,\gamma,\eta\rangle = 0,\label{abgest}
\end{equation}
where $\hat{\widetilde{a}}_p$ and $\hat{\widetilde{b}}_p$ are canonically transformed annihilation operators. Then one can calculate the corresponding matrix of propagators. The retarded and advanced propagators will be the same as the corollary of (\ref{bt02})--(\ref{alphapq}). But one will obtain different Kledysh popagators.

The new Keldysh propagator, calculated with the use of an excited state on top of (\ref{abgest}) has the following form:

\begin{eqnarray}
    \label{eq5}
    \tilde{G}^{K}(t,x;t^{\prime},x^{\prime}) \equiv -i\langle \Omega|\big[\tilde{\psi}(t,x),\bar{\tilde{\psi}}(t^{\prime},x^{\prime})\big]| \Omega\rangle = \nonumber    \\
    =-i\int \dfrac{dp}{2\pi}\int \dfrac{dq}{2\pi}\bigg[\tilde{\chi}_{p}(t,x)\bar{\tilde{\chi}}_{q}(t^{\prime},x^{\prime})\bigg(2\pi \delta(p-q)-2n^{\prime}_{qp}\bigg)+\tilde{\zeta}_{p}(t,x)\bar{\tilde{\zeta}}_{q}(t^{\prime},x^{\prime})\bigg(2\tilde{n}^{\prime}_{pq}-2\pi \delta(p-q)\bigg)\biggr] = \nonumber \\
    = G^{K}(t,x;t^{\prime},x^{\prime})+2i\int \dfrac{dp}{2\pi}\int \dfrac{dq}{2\pi}\bigg[n_{qp}\chi_{p}(t,x)\bar{\chi}_{q}(t^{\prime},x^{\prime})-\tilde{n}_{pq}\zeta_{p}(t,x)\bar{\zeta}_{q}(t^{\prime},x^{\prime}) - \nonumber \\
    -\kappa_{pq}\chi_{p}(t,x)\bar{\zeta}_{q}(t^{\prime},x^{\prime})-\kappa^{\dagger}_{pq}\zeta_{q}(t,x)\bar{\chi}_{p}(t^{\prime},x^{\prime})\bigg], \ \ \ \
\end{eqnarray}
where $G^{K}(t,x;t^{\prime},x^{\prime})$ is given by eq. (\ref{ft21}),  $\tilde{\psi}$ is a field operator rewritten in terms of $\tilde{\chi}$ and $\tilde{\zeta}$ and
\begin{equation}
    \label{03}
    \begin{cases}
    n^{\prime}_{pq} \equiv \langle \Omega |\hat{\tilde{a}}_{p}^{\dagger}\hat{\tilde{a}}_{q} |\Omega \rangle \\
       \tilde{n}^{\prime}_{pq} \equiv \langle \Omega |\hat{\tilde{b}}_{p}^{\dagger}\hat{\tilde{b}}_{q} |\Omega \rangle    \\
    \end{cases}.
\end{equation}
Also
\begin{equation}
    \label{equation1}
    n_{qp} \equiv \int_{}^{}\dfrac{dk_{1}}{2\pi} \ \gamma_{k_{1}p} \gamma^{*}_{k_{1}q}+\iint\dfrac{dk_{1}}{2\pi}\dfrac{dk_{2}}{2\pi}\bigg[n^{\prime}_{k_{2}k_{1}}\alpha_{k_{1}p}\alpha^{*}_{k_{2}q}-\tilde{n}^{\prime}_{k_{1}k_{2}}\gamma_{k_{1}p}\gamma^{*}_{k_{2}q}\bigg],
\end{equation}
\begin{equation}
    \label{equation2}
    \tilde{n}_{pq} \equiv \int_{}^{}\dfrac{dk_{1}}{2\pi} \ \beta_{k_{1}p} \beta^{*}_{k_{1}q}+\iint\dfrac{dk_{1}}{2\pi}\dfrac{dk_{2}}{2\pi}\bigg[\tilde{n}^{\prime}_{k_{1}k_{2}}\eta_{k_{1}p}\eta^{*}_{k_{2}q}-n^{\prime}_{k_{2}k_{1}}\beta_{k_{1}p}\beta^{*}_{k_{2}q}\bigg],
\end{equation}
and
\begin{equation}
    \label{equation3}
    \kappa_{pq} \equiv \int_{}^{}\dfrac{dk_{1}}{2\pi}\alpha_{k_{1}p} \beta^{*}_{k_{1}q}+\iint\dfrac{dk_{1}}{2\pi}\dfrac{dk_{2}}{2\pi}\bigg[\tilde{n}^{\prime}_{k_{1}k_{2}}\gamma_{k_{1}p}\eta^{*}_{k_{2}q}-n^{\prime}_{k_{2}k_{1}}\alpha_{k_{1}p}\beta^{*}_{k_{2}q}\bigg].
\end{equation}
On the other hand we know the exact form of the Keldysh propagator for a given distribution function $f$. Then, one should expect that for a state $|\Omega\rangle$ and for some choice of $\alpha$, $\beta$, $\gamma$ and $\eta$ there is a relation:

\begin{eqnarray}
    \label{eq8}
    G^{K}(t,x;t^{\prime},x^{\prime})+2i\int \dfrac{dp}{2\pi}\int \dfrac{dq}{2\pi}\bigg[n_{qp}\chi_{p}(t,x)\bar{\chi}_{q}(t^{\prime},x^{\prime})-\tilde{n}_{pq}\zeta_{p}(t,x)\bar{\zeta}_{q}(t^{\prime},x^{\prime}) - \nonumber \\
    -\kappa_{pq}\chi_{p}(t,x)\bar{\zeta}_{q}(t^{\prime},x^{\prime})-\kappa^{\dagger}_{pq}\zeta_{q}(t,x)\bar{\chi}_{p}(t^{\prime},x^{\prime})\bigg] \equiv \nonumber \\
    \equiv \int d\tau G^{R}(t,x;\tau,x^{\prime})f(\tau-t^{\prime})-\int d\tau f(t-\tau)G^{A}(\tau,x;t^{\prime},x^{\prime})+ \nonumber \\
    +\int dy \int d\tau_{1} d\tau_{2}G^{R}(t,x;\tau_{1},y)G^{A}(\tau_{2},y;t^{\prime},x^{\prime})\big[\Phi(\tau_{2},y)-\Phi(\tau_{1},y)\big]f(\tau_{1}-\tau_{2}) \ .
\end{eqnarray}
If we will solve this equation for $n$'s and $\kappa$'s, then that will establish the connection between the states in operator and functional formalisms.

\section{The behaviour of the previously found Keldysh propagators at past infinity}

At past infinity the exact Keldysh propagator $G^{K}$ which was found in the functional formalism has the following form:

\begin{equation}
    \label{ft20}
    G^{K}(v,u;v^{\prime},u^{\prime}) \approx \left[ {\begin{array}{cc}
   0&  \mathcal{P}\dfrac{T}{\sinh{\big[\sqrt{2}\pi T(u^{\prime}-u)\big]}} \\
    \mathcal{P}\dfrac{T}{\sinh{\big[\sqrt{2}\pi T(v^{\prime}-v)\big]}} & 0 \\
  \end{array} } \right],
\end{equation}
as follows directly from the Dyson--Schwinger equation,
if the background field is switched off $\Phi(v) \rightarrow 0$, when $v \rightarrow -\infty$, as we have assumed.

Let us see what happens at past infinity with the Keldysh propagator which was calculated in the operator formalism (\ref{ft21}). It is straightforward to see that its diagonal components behave as:

\begin{equation}
    \label{ft23}
    \begin{cases}
    G_{11}^{K}(v,u;v^{\prime},u^{\prime}) \approx \Phi(v)\ln\sqrt{(u-u^{\prime})a(v,v^{\prime})} \approx \Phi \ln(\Phi) \approx 0,  \\
    G_{22}^{K}(v,u;v^{\prime},u^{\prime}) \approx \Phi(v^{\prime})\ln\sqrt{(u-u^{\prime})a(v,v^{\prime})} \approx \Phi \ln(\Phi) \approx 0, \quad \text{when} \quad \Phi(v) \rightarrow 0
    \end{cases}
\end{equation}
One of the non--diagonal components is as follows:

\begin{eqnarray}
    \label{ft24}
    G_{12}^{K}(v^{\prime},u^{\prime};v,u)=-\dfrac{1}{\sqrt{2}\pi}\int_{0}^{+\infty}dp \  \sin{\bigg(p(u-u^{\prime})+a(v,v^{\prime})/p \bigg)}= \nonumber \\
    =-\dfrac{1}{\sqrt{2}\pi} \lim_{\delta^{v},\delta^{u} \rightarrow 0} Im \int_{0}^{+\infty}dp \  e^{-[\delta^{u}-i(u-u^{\prime})]p-[\delta^{v}-ia(v,v^{\prime})]/p }=\nonumber \\
    =-\dfrac{1}{\sqrt{2}\pi} \lim_{\delta^{u} \rightarrow 0} Im \dfrac{1}{\delta^{u}-i(u-u^{\prime})} = \dfrac{1}{\sqrt{2}\pi} \mathcal{P}\dfrac{1}{u^{\prime}-u},
\end{eqnarray}
while the other one is:

\begin{eqnarray}
    \label{ft25}
    G_{21}^{K}(v^{\prime},u^{\prime};v,u)=-\dfrac{1}{\sqrt{2}\pi}\dfrac{\Phi(v)\Phi(v^{\prime})}{2}\int_{0}^{+\infty}\dfrac{dp}{p^{2}} \  \sin{\bigg(p(u-u^{\prime})+a(v,v^{\prime})/p \bigg)}= \nonumber \\
    =-\dfrac{1}{\sqrt{2}\pi} \dfrac{\Phi(v)\Phi(v^{\prime})}{2} \lim_{\delta^{v},\delta^{u} \rightarrow 0} Im \int_{0}^{+\infty}\dfrac{dp}{p} \  e^{-[\delta^{u}-i(u-u^{\prime})]p-[\delta^{v}-ia(v,v^{\prime})]/p }= \bigg|p=\dfrac{1}{q}\bigg|=\nonumber \\
    =-\dfrac{1}{\sqrt{2}\pi} \dfrac{\Phi(v)\Phi(v^{\prime})}{2}\lim_{\delta^{v},\delta^{u} \rightarrow 0} Im \int_{0}^{+\infty}dq \  e^{-[\delta^{u}-i(u-u^{\prime})]/q-[\delta^{v}-ia(v,v^{\prime})]q }=\nonumber \\
    =-\dfrac{1}{\sqrt{2}\pi} \dfrac{\Phi(v)\Phi(v^{\prime})}{2} \lim_{\delta^{v} \rightarrow 0} Im \dfrac{1}{\delta^{v}-ia(v,v^{\prime})}=-\dfrac{1}{\sqrt{2}\pi} \dfrac{\Phi(v)\Phi(v^{\prime})}{2} \mathcal{P}\dfrac{1}{a(v,v^{\prime})}.
\end{eqnarray}
In all, for such a background field, which obeys the condition $\Phi(v) \rightarrow 0$, as $v\to - \infty$, we find that in the limit $v,v^{\prime} \rightarrow -\infty$:

\begin{equation}
    \label{ft26}
    G^{K}(v,u;v^{\prime},u^{\prime}) \approx \left[ {\begin{array}{cc}
   0&  \dfrac{1}{\sqrt{2}\pi}\mathcal{P}\dfrac{1}{u^{\prime}-u} \\
    -\dfrac{1}{\sqrt{2}\pi} \dfrac{\Phi(v)\Phi(v^{\prime})}{2} \mathcal{P}\dfrac{1}{a(v,v^{\prime})} & 0 \\
  \end{array} } \right].
\end{equation}
Thus, the Keldysh propagator found in the operator formalism does not necessarily reduce to its vacuum form at the past infinity, due to turning of the background field. That is quite surprising, but is due to the fact that the modes in eq. (\ref{eq:132}) contain integrals over $v$, which is some sort of a memory encoded into their basis. Let us examine the situation in grater details.

If we assume that $\Phi(v) = 0$ when $v<v_0$ for some $v_0$ and $v,v'$ in $a(v,v')$ are both smaller than $v_0$, then eq. (\ref{ft26}) reduces to (\ref{ft20}) for $T=0$. I.e. in this case the background field does not affect the propagator at past infinity. However, consider e.g. such a class of background fields which behave as:

\begin{equation}
    \label{ft27}
    \Phi(v)\approx Ce^{\alpha v}, \quad \text{when} \quad v \rightarrow -\infty.
\end{equation}
Then, because

$$a(v,v^{\prime})=\dfrac{1}{2}\int_{v^{\prime}}^{v} dy \, \Phi^{2}(y) = \dfrac{C^{2}}{4\alpha}\bigg(e^{2\alpha v}-e^{2\alpha v^{\prime}}\bigg)=\dfrac{C^{2}}{2\alpha}e^{\alpha(v+v^{\prime})}\sinh{\big[\alpha(v-v^{\prime})\big]},$$
the $G^K_{21}$ component of (\ref{ft26}) behaves as

\begin{equation}
    \label{ft28}
    -\dfrac{1}{\sqrt{2}\pi} \dfrac{\Phi(v)\Phi(v^{\prime})}{2} \mathcal{P}\dfrac{1}{a(v,v^{\prime})}\approx \mathcal{P}\dfrac{T_{0}}{\sinh{\big[\sqrt{2}\pi  T_{0}(v^{\prime}-v)\big]}},
\end{equation}
where $T_{0}\equiv \dfrac{\alpha}{\sqrt{2}\pi}$. As a result the Keldysh propagator under consideration behaves as

\begin{equation}
    \label{ft29}
    G^{K}(v,u;v^{\prime},u^{\prime}) \approx \left[ {\begin{array}{cc}
   0&  \dfrac{1}{\sqrt{2}\pi}\mathcal{P}\dfrac{1}{u^{\prime}-u} \\
    \mathcal{P}\dfrac{T_{0}}{\sinh{\big[\sqrt{2}\pi T_{0}(v^{\prime}-v)\big]}} & 0 \\
  \end{array} } \right], \quad {\rm when} \quad t \to - \infty.
\end{equation}
So, if we assume that the background field is turned on exponentially at past infinity, then it means that initially, the right handed fermions are at the ground state, while the left handed fermions are at the thermal equilibrium with some effective temperature $T_{0}$ set up by the way one turns on the background field. Moreover, with adiabatic turning on of the background field $\Phi$ the integral $a(v,v')$ can be non--trivial, which results in a non--trivial $G^K$ at past infinity due to the specifics of the process of turning on the background field. This observation is also consistent with expression for stress energy tensor. One can check that Schwarzian derivative in that limit is proportional to $\sim\alpha^2\sim T_0^2$ as it should be for chiral massless fermions at finite temperature. Thus, in the operator formalism, as opposed to the functional one, and the corresponding state there is no symmetry between the left and right moving fermions.

\section{Conclusions}

We have found that the exact retarded and advanced propagators are the same in the functional and operator formalisms. That is essentially just a consistency check for the validity of our calculations.

Then, we have found that physical observables calculated with the use of these two formalisms have different values, which is related to the fact that there are physically distinct states in the calculations in these two approaches. However, there are certain quantities that have the same values independently of the approach.
Namely, the form of the scalar current is the same in both situations for the large and slowly changing values of the background field $\Phi$. Subleading terms in the current of course are state dependent.

Furthermore, the form of the scalar current can be traced to the Feynman in--out effective action calculated in the approximation (\ref{eq:100}). In fact, if one were taking the T--ordered Feynman functional integral:

$$
e^{i \, S_{eff}[\Phi]} = \int D\psi \, D\bar{\psi} \, e^{i \, S[\psi, \bar{\psi}, \Phi]},
$$
in the limit (\ref{eq:100}) the result would have been that $S_{eff}[\Phi] \approx \int d^2x \, V_{eff}[\Phi]$. It is a textbook calculation to see that $V_{eff}$ is exactly such that

$$
\langle\bar{\psi}(t,x-\delta)\psi(t,x)\rangle \approx \frac{\partial V_{eff}(\Phi)}{\partial \Phi} \approx - \dfrac{\Phi(v)}{\pi}\ln\bigg[\Phi(v)\, \delta \bigg].
$$
There are several points which are worth stressing here. First, these observations mean that for the properly chosen modes and/or propagators (i.e. such that they have proper UV Hadamard behaviour) the form of the current in the limit (\ref{eq:100}) is universal and state independent. This seems to be natural, because in this limit the field is very strong and, hence, is not sensitive, at the leading order, to the properties of the low laying ground state.

Second, the Feynman effective action does not have any imaginary contribution, contrary to what is happening e.g. in the strong electric field. Moreover, unlike e.g. the case of the strong electric field the effective action in this case is analytic function on the complex cut $\Phi$--plane. Which means that there is no tunneling of fermions in the background scalar field. That seems to signal that there is no particle creation in the situation that we are discussing here.

However, as we observe in the subsection 3.4 the situation is not that simple. Because there we encounter a non--trivial fermion flux in the operator formalism. The flux is given by:

\begin{equation}
   \langle T^{01} \rangle_{reg} = -\dfrac{1}{48\pi}\{\bar{a}(v),v\},
\end{equation}
where

$$
\bar{a}(v)=\dfrac{1}{m^{2}}\int_{}^{v}dy \, \Phi^{2}(y),
$$
and $\{f(z),z\}$ is the Schwarzian derivative:
\begin{equation}
        \{f(z),z\}=\dfrac{f^{\prime \prime \prime}(z)}{f^{\prime}(z)}-\dfrac{3}{2}\bigg(\dfrac{f^{\prime \prime}(z)}{f^{\prime}(z)}\bigg)^{2} \nonumber .
\end{equation}
This fact definitely signals fermion creation. At the same time in the functional formalism we find that the flux is zero. The situation is similar to the one on the black hole background over the Bouleward and Hartle--Hawking vs. Unruh states \cite{Candelas:1980zt}.

Furthermore, the QFT in the background field $\Phi$ as given by eq. (\ref{eq:131}) is similar to the QFT in the presence of the non--ideal mirror \cite{Akhmedov:2017hbj}, \cite{Astrahantsev:2018sxg}. Such a mirror is transparent for high energy modes unlike the ideal one \cite{Davies:1977yv}, \cite{Davies:1976hi}, which reflects waves of any energy. Furthermore, similarly to the case under consideration, the Schwarzian type of stress--energy flux naturally appears in the case of moving mirrors \cite{Davies:1977yv}, \cite{Davies:1976hi}.

Third, we have seen that $\Phi(t-x)$ solves the classical equations of motion $\partial^2 \phi = 0$. However, on the quantum level, zero point fluctuations create an effective potential and the expectation value of the eq. (\ref{M:8}) reduces to:

$$
\partial^2 \phi \approx - \lambda^2 \,  \dfrac{\phi}{\pi}\ln\bigg[\phi\, \delta \bigg],
$$
for the large and slowly changing values of $\phi$.

Now $\Phi(t-x)$ is not a solution anymore of the effective equations of motion. The correct solution should describe rolling of the field $\phi$ down the to the minimum of the effective potential. Actually to describe the rolling one probably needs a bit more expanded form of the effective action. The latter depends on the choice of the quantum ground state.

During this rolling there will probably be quantum loop amplification of the tree-level particle flux due to the change of the anomalous quantum averages and of the level population for fermions and scalar. But that can be seen only in the loops when the field $\phi$ is made dynamical. This is at least what we already have seen in the de Sitter space background \cite{Akhmedov:2019cfd}--\cite{Akhmedov:2011pj}, in the black hole collapse background \cite{Akhmedov:2015xwa}, in the strong electric field backgrounds \cite{Akhmedov:2014doa}, \cite{Akhmedov:2014hfa}, in the case of moving mirrors \cite{Akhmedov:2017hbj}, \cite{Alexeev:2017ath} (see also the case of non--linear quantum mechanics with time dependent frequency \cite{Trunin:2018egi}).

\section{Acknowledgments}

We would like to thank C.Schubert, V.Losyakov, F.Popov, D.Trunin, E.Lanina and G.Jorjadze for useful discussions. AET would like to thank Hermann Nicolai and Stefan Theisen for the hospitality at the Albert Einstein Institute, Golm, where the work on this project was completed. This work was supported by the Russian Ministry of Science and Education, project number 3.9904.2017/BasePart. The work of ETA was supported by the grant from the Foundation for the Advancement of Theoretical Physics and Mathematics “BASIS” and by RFBR grant 18-01-00460 A.

%\newpage

\appendix

\section{Free fermions without background field at finite temperature} \label{A}

To make our paper selfcontained in this Appendix we discuss the properties of the propagators in the free massless 2D fermionic theory at finite temperature.
The action of the theory has the form:

\begin{equation}
    \label{eq:72}
     S[\psi, \bar{\psi}]= \int d^2x \, \bar{\psi}(x,t)i \slashed{\partial}\psi(x,t) \ .
\end{equation}
The equations of motion are:

\begin{equation}
    \label{eq:73}
  \left[ {\begin{array}{cc}
   0 & i\bigg(\partial_{t}+\partial_{x}\bigg) \\
    i\bigg(\partial_{t}-\partial_{x}\bigg) & 0 \\
  \end{array} } \right]\begin{bmatrix}\psi_{R} \\ \psi_{L} \end{bmatrix}=0 \ .
\end{equation}
We quantize the theory in a box of length
$L$ with periodic boundary conditions and then take $L\to \infty$. The free field
solutions to the Dirac equation are given by
\begin{equation}
    \label{eq:74}
    \hat{\psi}_{\eta}=\sum_{p>0}^{} \sqrt{\dfrac{1}{L}}\bigg(\hat{a}_{\eta p}e^{-ip(t+\eta x)}+\hat{b}^{\dagger}_{\eta p}e^{ip(t+\eta x)}\bigg) \ ,
\end{equation}
where $\eta=+$ for the right moving component and $\eta=-$ for the left moving component of the spinor.
The creation and annihilation operators satisfy the canonical anti--commutation relations $ \{\hat{a}_{\eta p},\hat{a}^{\dagger}_{\eta p^{\prime}}\}= \delta_{pp^{\prime}}$ and $\{\hat{b}_{\eta p},\hat{b}^{\dagger}_{\eta p^{\prime}}\}=\delta_{pp^{\prime}}$.
The free Hamiltonian is:

\begin{equation}
    \label{eq:75}
    \hat{H}=\int dx \bigg(\hat{\psi}_{R}^{\dagger}i\partial_{x}\hat{\psi}_{R}-\hat{\psi}_{L}^{\dagger}i\partial_{x}\hat{\psi}_{L}\bigg)=\sum_{p>0}^{}\sum_{\eta=R,L}^{} p\bigg(\hat{a}^{\dagger}_{\eta p}\hat{a}_{\eta p}+\hat{b}^{\dagger}_{\eta p}\hat{b}_{\eta p}\bigg) \ .
\end{equation}
The expectation value of an operator $\hat{A}$ is defined as:

\begin{equation}
    \label{eq:76}
    \langle \hat{A} \rangle=Tr\big(\hat{\rho}\hat{A}\big) \ ,
\end{equation}
where the trace is taken in Fock space and $T$ is the temperature. For the thermal state the density matrix operator is as follows:

\begin{equation}
    \label{eq:77}
    \hat{\rho}=\dfrac{e^{-\beta\hat{H}}}{Tr\big(e^{-\beta\hat{H}}\big)}=\dfrac{1}{Z}e^{-\beta\hat{H}}, \, \quad \beta = \frac{1}{T}.
\end{equation}
$Z$ is the canonical partition function.
To calculate the Green functions in the thermal state we need to use the expectation value of the occupation number operator:

\begin{equation}
    \label{eq:78}
    n_{p} \delta(p-q) = \langle \hat{a}^{\dagger}_{\eta p}\hat{a}_{\eta q} \rangle = \langle \hat{b}^{\dagger}_{\eta p}\hat{b}_{\eta q} \rangle = \dfrac{\delta(p-q)}{e^{\beta p}+1} \ .
\end{equation}
Then the Wightman propagator as a matrix in spinor indexes has the following form:

\begin{equation}
    \label{eq:79}
     G_{0}=-i\langle \hat{\psi}(t,x)\hat{\psi}^{\dagger}(t^{\prime},x^{\prime}) \rangle=\left[ {\begin{array}{cc}
   0 & -i\langle \hat{\psi}_{R}(t,x)\hat{\psi}_{R}^{\dagger}(t^{\prime},x^{\prime})\rangle \\
    -i\langle \hat{\psi}_{L}(t,x)\hat{\psi}_{L}^{\dagger}(t^{\prime},x^{\prime})\rangle & 0 \\
  \end{array} } \right] \ .
\end{equation}
The Feynman propagator is:

\begin{equation}
    \label{eq:80}
    G^{T}(t,x;t^{\prime},x^{\prime})=\theta(t-t^{\prime})G^{>}+\theta(t^{\prime}-t)G^{<},
\end{equation}
where $G^>$ and $G^<$ are defined in eq.(\ref{eq:10}); the retarded Green function is:

\begin{equation}
    \label{eq:81}
    G^{R}(t,x;t^{\prime},x^{\prime})=\theta(t-t^{\prime})\bigg(G^{>}-G^{<}\bigg);
\end{equation}
the advanced Green function is:

\begin{equation}
    \label{eq:82}
    G^{A}(t,x;t^{\prime},x^{\prime})=-\theta(t^{\prime}-t)\bigg(G^{>}-G^{<}\bigg);
\end{equation}
and the Keldysh propagator is:

\begin{equation}
    \label{eq:83}
    G^{K}(t,x;t^{\prime},x^{\prime})=G^{>}+G^{<} \ .
\end{equation}
Then the explicit form of the retarded propagator for the right moving fermions is:

\begin{equation}
    \label{eq:84}
    G^{R}_{r}=-i\theta(t-t^{\prime})\sum_{p,p^{\prime}>0}^{}\dfrac{1}{L}\bigg([\hat{a}_{R p},\hat{a}^{\dagger}_{R p^{\prime}}]_{+}e^{-ip(t+x-(t^{\prime}+x^{\prime}))}+[\hat{b}^{\dagger}_{R p},\hat{b}_{R p^{\prime}}]_{+}e^{ip(t+x-(t^{\prime}+x^{\prime}))}\bigg)=
\end{equation}
$$=-i\theta(t-t^{\prime})\sum_{p>0}^{}\dfrac{1}{L}\bigg(e^{-ip(t+x-(t^{\prime}+x^{\prime}))}+e^{ip(t+x-(t^{\prime}+x^{\prime}))}\bigg)=-\dfrac{i}{\sqrt{2}}\theta(v-v^{\prime})\delta(u-u^{\prime}) \ .$$
Similarly, for the left moving fermions we have:

\begin{equation}
    \label{eq:85}
    G^{R}_{l}=-\dfrac{i}{\sqrt{2}}\theta(u-u^{\prime})\delta(v-v^{\prime}) \ .
\end{equation}
Thus, for the retarded Green function we obtain the expression from eq. (\ref{eq:45}).

Let us calculate now the Keldysh propagator for the right moving fermions:

\begin{equation}
    \label{eq:86}
    G^{K}_{r}(t,x;t^{\prime},x^{\prime})=-i\sum_{p,p^{\prime}>0}^{}\dfrac{1}{L}\bigg([\hat{a}_{R p},\hat{a}^{\dagger}_{R p^{\prime}}]_{-}e^{-ip(t+x-(t^{\prime}+x^{\prime}))}+[\hat{b}^{\dagger}_{R p},\hat{b}_{R p^{\prime}}]_{-}e^{ip(t+x-(t^{\prime}+x^{\prime}))}\bigg)=
\end{equation}
$$=-i\sum_{p>0}^{}\dfrac{1-2n_{p}}{L}\bigg(e^{-ip(t+x-(t^{\prime}+x^{\prime}))}-e^{ip(t+x-(t^{\prime}+x^{\prime}))}\bigg)=-i\int_{-\infty}^{+\infty}\dfrac{dp}{2\pi}\tanh{\bigg(\dfrac{p}{2T}\bigg)}e^{-ip(t-t^{\prime}+x-x^{\prime})} \ .$$
The Fourier transformation of the Green functions \eqref{eq:84}, \eqref{eq:85}, \eqref{eq:86} and of the advanced propagator, as defined in \eqref{eq:125}, is as follows:

\begin{equation}
\label{eq:B1}
    \begin{cases}
    G^{R}(\epsilon,x;\epsilon^{\prime},x^{\prime})=-2\pi i\delta(\epsilon+\epsilon^{\prime})\theta(x^{\prime}-x)e^{-i\epsilon(x-x^{\prime})} \\
    G^{A}(\epsilon,x;\epsilon^{\prime},x^{\prime})=2\pi i\delta(\epsilon+\epsilon^{\prime})\theta(x-x^{\prime})e^{-i\epsilon(x-x^{\prime})} \\
    G^{K}(\epsilon,x;\epsilon^{\prime},x^{\prime})=-2\pi i\delta(\epsilon+\epsilon^{\prime})\tanh{\bigg(\dfrac{\epsilon}{2T}\bigg)}e^{-i\epsilon(x-x^{\prime})} \ .
    \end{cases}
\end{equation}
From \eqref{eq:B1} it follows, that there is a relation between $G^{R/A}$ and $G^{K}$:

\begin{equation}
    \label{eq:B2}
    G^{K}(\epsilon,x;\epsilon^{\prime},x^{\prime})=\tanh{\bigg(\dfrac{\epsilon}{2T}\bigg)} \bigg[G^{R}(\epsilon,x;\epsilon^{\prime},x^{\prime})-G^{A}(\epsilon,x;\epsilon^{\prime},x^{\prime})\bigg] \ ,
\end{equation}
which is the essence of the so called Fluctuation-Dissipation Theorem \cite{LL}.

Let us find now the inverse Fourier transformation of the distribution function $F(\epsilon) = \tanh \frac{\epsilon}{2T}$:

\begin{equation}
    \label{eq:88}
    f(\tau)=\int\limits_{-\infty}^{+\infty} \dfrac{d\epsilon}{2\pi}e^{-i\epsilon \tau} \, \tanh\dfrac{\epsilon}{2T} = -\int\limits_{-\infty}^{+\infty} \dfrac{d\epsilon}{2\pi} \, e^{-i\epsilon \tau}\bigg(1-2\dfrac{1}{e^{\beta \epsilon}+1}\bigg)=
\end{equation}
$$=-\delta(\tau)+\lim_{\delta \to 0}\dfrac{1}{\pi}\int_{-\infty}^{+\infty} d\epsilon \,  e^{-\epsilon(\delta+i\tau)} \, \dfrac{1}{e^{\beta \epsilon}+1}=-\delta(\tau)-\lim_{\delta \to 0}\dfrac{iT}{\sinh{\big(\pi T(\tau-i\delta)\big)}} \ .$$
Then, if $\tau \neq 0$, we have that

$$
f(\tau)=-\dfrac{iT}{\sinh{\big(\pi T \tau\big)}},
$$
while if $\tau \rightarrow{0}$:

\begin{equation}
    \label{eq:89}
    f(\tau)=-\delta(\tau)-\dfrac{i}{\pi}\dfrac{1}{\tau-i\delta}=-\delta(\tau)-\dfrac{i}{\pi}\dfrac{\tau+i\delta}{\tau^{2}+\delta^{2}}=-\dfrac{i}{\pi}\lim_{\delta \to 0}\dfrac{\tau}{\tau^{2}+\delta^{2}}\ .
\end{equation}
In all, one can write that:

\begin{equation}
    \label{eq:90}
    f(\tau)=-\mathcal{P}\dfrac{iT}{\sinh{\big(\pi T \tau\big)}} \ .
\end{equation}

\section{Calculation of the retarded and advanced propagators in the functional formalism} \label{B}

In components the equation \eqref{eq:25} can be written as:

\begin{equation}
    \begin{cases}{}
 -\Phi(v)G^{R}_{11}+i\sqrt{2}\partial_{u}G^{R}_{21}=\delta(u-u^{\prime})\delta(v-v^{\prime}) \ , \\
i\sqrt{2}\partial_{v}G^{R}_{12}-\Phi(v)G^{R}_{22}=\delta(u-u^{\prime})\delta(v-v^{\prime}) \ , \\
-\Phi(v)G^{R}_{12}+i\sqrt{2}\partial_{u}G^{R}_{22}=0 \ , \\
i\sqrt{2}\partial_{v}G^{R}_{11}-\Phi(v)G^{R}_{21}=0 \ . \label{eq:31}
\end{cases}
\end{equation}
From here we get that:

\begin{equation}
    \label{eq:32}
    \bigg(2\partial_{v}\partial_{u}+\Phi^{2}(v)\bigg)G^{R}_{11}=-\Phi(v^{\prime})\delta(u-u^{\prime})\delta(v-v^{\prime}),
\end{equation}
and
\begin{equation}
    \label{eq:33}
    \bigg(2\partial_{v}\partial_{u}-2\big[\ln{\Phi(v)}]^{\prime}\partial_{u}+\Phi^{2}(v)\bigg)G^{R}_{22}=-\Phi(v^{\prime})\delta(u-u^{\prime})\delta(v-v^{\prime}).
\end{equation}
First we solve \eqref{eq:32} and \eqref{eq:33}, and then we find $G^{R}_{12}$ and $G^{R}_{21}$ from \eqref{eq:31}.
Since the potential $\Phi$ dependents only on $v$, we can represent $G^{R}_{11}$ as:
\begin{equation}
    \label{eq:34}
    G^{R}_{11}(u,v)=-\Phi(v^{\prime})\int \dfrac{dp}{2\pi}e^{ip(u-u^{\prime})}\dfrac{G^{R}_{11}(p,v)}{2ip} \ .
\end{equation}
After that eq. \eqref{eq:32} acquires the form:
\begin{equation}
    \label{eq:35}
    \bigg(\partial_{v}+\dfrac{\Phi^{2}(v)}{2ip}\bigg)G^{R}_{11}(p,v)=\delta(v-v^{\prime}) \ .
\end{equation}
A solution of this equation is as follows:

\begin{equation}
    \label{eq:36}
    G^{R}_{11}(p,v)=e^{-\dfrac{\int_{v^{\prime}}^{v}\Phi^{2}(v)}{2ip}}\theta(v-v^{\prime}) \ .
\end{equation}
For convenience we introduce the following notation:

\begin{equation}
    \label{eq:37}
    a(v,v') = \frac12 \int_{v'}^{v}dy \, \Phi^{2}(y) \geq 0.
\end{equation}
Then, using \cite{1}, we obtain that:

\begin{eqnarray}
    \label{eq:38}
    G^{R}_{11}(u,v)=-\Phi(v^{\prime})\theta(v-v^{\prime})\int \dfrac{dp}{2\pi}\dfrac{1}{2ip}e^{ip(u-u^{\prime})}e^{-\dfrac{a(v,v^{\prime})}{ip}}=
\end{eqnarray}
$$ =-\Phi(v^{\prime}) \, \theta(v-v^{\prime}) \, \int_{0}^{+\infty} \dfrac{dp}{2\pi} \, \dfrac{2i}{2ip} \, \sin\bigg(p(u-u^{\prime}) + \dfrac{a(v,v^{\prime})}{p}\bigg),$$
where

\begin{equation}
\label{eq:39}
  \int_{0}^{+\infty} \dfrac{dp}{2\pi} \, \dfrac{1}{p} \, \sin\bigg(p(u-u^{\prime}) + \dfrac{a(v,v^{\prime})}{p}\bigg) =
  \begin{cases}
    \dfrac{1}{2}J_{0}\bigg(2\sqrt{a(v,v^{\prime})(u-u^{\prime})}\bigg), & \text{for $u-u^{\prime} > 0$} \\
    \dfrac{1}{4}, & \text{for $u-u^{\prime} = 0$ } \\
    0, & \text{for $u-u^{\prime} < 0$}.
  \end{cases}
\end{equation}
Here $J_0(x)$ is the Bessel function.

In compact form $G^{R}_{11}$ can be written as:

\begin{equation}
    \label{eq:40}
    G^{R}_{11}(u,v)=-\dfrac{1}{2}\Phi(v^{\prime})\theta(v-v^{\prime})\theta(u-u^{\prime})J_{0}\bigg(2\sqrt{a(v,v^{\prime})(u-u^{\prime})}\bigg) \ .
\end{equation}
Repeating the same steps as above, we find that:
\begin{eqnarray}
    \label{eq:41}
    G^{R}_{22}(u,v)=\exp\bigg(\int_{v^{\prime}}^{v}\big[\ln \Phi(v)\big]^{\prime}\bigg)G^{R}_{11}(u,v)=\dfrac{\Phi(v)}{\Phi(v^{\prime})}G^{R}_{11}= \nonumber \\
    =-\dfrac{1}{2}\Phi(v)\theta(v-v^{\prime})\theta(u-u^{\prime})J_{0}\bigg(2\sqrt{a(v,v^{\prime})(u-u^{\prime})}\bigg) \ .
\end{eqnarray}
From \eqref{eq:31} it follows that:

\begin{equation}
    \label{eq:42}
    G^{R}_{12}=\dfrac{i\sqrt{2}}{\Phi(v)}\partial_{u}G^{R}_{22}=
\end{equation}
$$=-\dfrac{i}{\sqrt{2}}\theta(v-v^{\prime})\delta(u-u^{\prime})+\dfrac{i}{\sqrt{2}}\theta(v-v^{\prime})\theta(u-u^{\prime})\sqrt{\dfrac{a(v,v^{\prime})}{u-u^{\prime}}}J_{1}\bigg(2\sqrt{a(v,v^{\prime})(u-u^{\prime})}\bigg) \ ,$$
where we have used the following properties of the Bessel functions:

\begin{equation}
    \label{eq:43}
    J_{0}(0)=1 \ , \quad \quad J^{\prime}_{0}(x)=-J_{1}(x) \ .
\end{equation}
Then using again the system equations \eqref{eq:31}, we obtain that:

\begin{equation}
    \label{eq:44}
    G^{R}_{21}=\dfrac{i\sqrt{2}}{\Phi(v)}\partial_{v}G^{R}_{11}=
\end{equation}
$$=-\dfrac{i}{\sqrt{2}}\delta(v-v^{\prime})\theta(u-u^{\prime})+\dfrac{i}{\sqrt{2}}\Phi(v^{\prime})\theta(v-v^{\prime})\theta(u-u^{\prime})\sqrt{\dfrac{u-u^{\prime}}{a(v,v^{\prime})}}\dfrac{\Phi(v)}{2}J_{1}\bigg(2\sqrt{a(v,v^{\prime})(u-u^{\prime})}\bigg) \ .$$
Putting all the results of this Appendix together we obtain eq. (\ref{eq:q146}).

\section{Calculation of the anomalous contribution to the Keldysh propagator in the coincidence limit in the functional formalism}
\label{C}

For the further steps it is convenient to write $G^{R/A}$ from (\ref{eq:q146}) as:

\begin{equation}
    \label{eq:RM}
    \begin{cases}
     G^{R}=G_{0}^{R}+\theta(v-v^{\prime})\theta(u-u^{\prime})D^{R} \ , \\
      G^{A}=G_{0}^{A}+\theta(v^{\prime}-v)\theta(u^{\prime}-u)D^{A} \ ,
    \end{cases}
\end{equation}
where the exact form of $D^{R/A}$ follows from \eqref{eq:q146}. Then, using the last equation the product \\ $G^{R}(x,t;y,\tau_{1}) \, G^{A}(y,\tau_{2};x,t)$ can be written as:

\begin{eqnarray}
    \label{eq:64}
    G^{R}(x,t;y,\tau_{1})G^{A}(y,\tau_{2};x,t)=\nonumber \\
    =\bigg(G^{R}_{0}(x,t;y,\tau_{1})+\theta(t-x-\tau_{1}+y)\theta(t+x-\tau_{1}-y)D^{R}(x,t;y,\tau_{1})\bigg)\times \nonumber \\
    \times\bigg(G_{0}^{A}(y,\tau_{2};x,t)+\theta(t-x-\tau_{2}+y)\theta(t+x-\tau_{2}-y)D^{A}(y,\tau_{2};x,t)\bigg)=\nonumber \\
    =\bigg(G^{R}G^{A}\bigg)^{(1)}+\bigg(G^{R}G^{A}\bigg)^{(2)}+\bigg(G^{R}G^{A}\bigg)^{(3)} \ ,
\end{eqnarray}
where, we denote by $\bigg(G^{R}G^{A}\bigg)^{(i)}, i=1,2,3$ the contributions that have the following structures:

$$\bigg(G^{R}G^{A}\bigg)^{(1)} \sim \theta(...)\theta(...)\delta(...)\delta(...) \ ,$$
$$\bigg(G^{R}G^{A}\bigg)^{(2)} \sim \theta(...)\theta(...)\theta(...)\theta(...) \ ,$$
$$\bigg(G^{R}G^{A}\bigg)^{(3)} \sim \theta(...)\theta(...)\theta(...)\delta(...) \ .$$
As a result one can divide contributions to $G^{K}_{an}$ as follows:

\begin{equation}
    \label{eq:160}
    G^{K}_{an} = G^{K}_{an1} + G^{K}_{an2} + G^{K}_{an3},
\end{equation}
where

\begin{eqnarray}
    \label{eq:161}
    G^{K}_{an1} \equiv \int dy \int d\tau_{1} d\tau_{2}\bigg(G^{R}(x,t;y,\tau_{1})G^{A}(y,\tau_{2};x,t)\bigg)^{(1)}\big[\Phi(\tau_{2}-y)-\Phi(\tau_{1}-y)\big]f(\tau_{1}-\tau_{2}) \ , \nonumber \\
    G^{K}_{an2} \equiv \int dy \int d\tau_{1} d\tau_{2}\bigg(G^{R}(x,t;y,\tau_{1})G^{A}(y,\tau_{2};x,t)\bigg)^{(2)}\big[\Phi(\tau_{2}-y)-\Phi(\tau_{1}-y)\big]f(\tau_{1}-\tau_{2}) \ , \nonumber \\
    G^{K}_{an3} \equiv \int dy \int d\tau_{1} d\tau_{2}\bigg(G^{R}(x,t;y,\tau_{1})G^{A}(y,\tau_{2};x,t)\bigg)^{(3)}\big[\Phi(\tau_{2}-y)-\Phi(\tau_{1}-y)\big]f(\tau_{1}-\tau_{2}) \ .
\end{eqnarray}
In \eqref{eq:161} one can make the following change:

\begin{eqnarray}
    \label{eq: ,161}
    \bigg(G^{R}(x,t;y,\tau_{1})G^{A}(y,\tau_{2};x,t)\bigg)^{(i)} \rightarrow{ } \nonumber \\ \rightarrow{} \dfrac{1}{2}\bigg[\bigg(G^{R}(x,t;y,\tau_{1})G^{A}(y,\tau_{2};x,t)\bigg)^{(i)}+\bigg(G^{R}(x,t;y,\tau_{2})G^{A}(y,\tau_{1};x,t)\bigg)^{(i)}\bigg],
\end{eqnarray}
since $\big[\Phi(\tau_{2}-y)-\Phi(\tau_{1}-y)\big]f(\tau_{2}-\tau_{1})$ is symmetric under the exchange of $\tau_{1} \leftrightarrow \tau_{2}$. Then, from \eqref{eq:45} and \eqref{eq:46} it immidiatly follows that all spinor components of $G^{K}_{an1}$ are zero.

To calculate $G^{K}_{an2}$ consider

\begin{equation}
    \label{eq:65}
    \bigg(G^{R}G^{A}\bigg)^{(2)}=\theta(t-x-\tau_{1}+y)\theta(t+x-\tau_{1}-y)\theta(t-x-\tau_{2}+y)\theta(t+x-\tau_{2}-y)D^{R}(x,t;y,\tau_{1})D^{A}(y,\tau_{2};x,t) \ .
\end{equation}
The product of the four Heaviside functions restricts the arguments to the region:
\begin{equation}
    \label{eq:66}
      \begin{cases}
    t>\tau_{1}\\
    t>\tau_{2}\\
    (t-\tau_{1})+x>y>-(t-\tau_{1})+x\\
    (t-\tau_{2})+x>y>-(t-\tau_{2})+x \ ,
    \end{cases}
\end{equation}
hence, after some manipulations this part of the Keldysh propagator can be written in the following form:

\begin{eqnarray}
    \label{eq:69}
     G^{K}_{an2}(x,t;x,t)= \nonumber \\
     =\int_{0}^{+\infty} d\tau_{1}\int_{0}^{+\infty} d\tau_{2} \int_{-min(\tau_{1},\tau_{2})}^{min(\tau_{1},\tau_{2})}dy \,  D^{R}(x,t;y+x,t-\tau_{1})D^{A}(y+x,t-\tau_{2};x,t)\times  \nonumber \\
     \times \bigg[\Phi\bigg(v-\dfrac{
     \tau_{2}+y}{\sqrt{2}}\bigg)-\Phi \bigg(v-\dfrac{
     \tau_{1}+y}{\sqrt{2}}\bigg)\bigg]f(\tau_{2}-\tau_{1}) \ .
\end{eqnarray}
Diagonal components of this spinor matrix have the following form:

\begin{eqnarray}
    \label{eq:163}
   [G^K_{an2}]_{11}(x,t;x,t) = \nonumber \\
   = -2 \sqrt{2} \int_{0}^{+\infty} d\tau_1\int_{0}^{+\infty} d\tau_2 \int_{-min(\tau_1,\tau_2)}^{min(\tau_1,\tau_2)} dy \ \bigg\{\Phi\left[v-(\tau_2+y)\right] - \Phi\left[v-(\tau_1+y)\right]\bigg\}
   \times  \nonumber \\ \times
   f\left[\sqrt{2}(\tau_1 - \tau_2)\right] \, \Bigg\{\frac14 \, \Phi(v) \,
   \Phi\left[v-(\tau_1 + y)\right] \, J_0\left[z(\tau_1,y)\right] \, J_0\left[z(\tau_2,y)\right] + \nonumber \\
   + \frac12 \, \Phi(v) \, \Phi\left[v-(\tau_2+y)\right]  \, J_{1}\left[z(\tau_1,y)\right] J_1\left[z(\tau_2,y)\right] \, \frac{z(\tau_1,y)}{z(\tau_2,y)} \,  \frac{\big(\tau_2-y\big)}{\big(\tau_1-y\big)} \Bigg\},
\end{eqnarray}
and
\begin{eqnarray}
    \label{eq:164}
   [G^K_{an2}]_{22}(x,t;x,t) = \nonumber \\
   =- 2 \sqrt{2} \int_0^{+\infty} d\tau_1\int_0^{+\infty} d\tau_2 \int_{-min(\tau_1,\tau_2)}^{min(\tau_1,\tau_2)} dy \, \bigg\{\Phi\left[v-(\tau_2 + y)\right] - \Phi\left[v-(\tau_1 + y)\right]\bigg\} \times \nonumber \\
   \times  f\left[\sqrt{2}(\tau_1 - \tau_2)\right] \, \Bigg\{\frac14 \, \Phi(v) \, \Phi\left[v-(\tau_2 + y)\right] \, J_0\left[z(\tau_1,y)\right] \, J_0\left[z(\tau_2,y)\right] + \nonumber \\
   + \frac12 \, \Phi(v) \, \Phi\left[v-(\tau_1+y)\right] \, J_1\left[z(\tau_1,y)\right] \, J_1\left[z(\tau_2,y)\right] \, \frac{z(\tau_2,y)}{z(\tau_1,y)} \, \frac{\big(\tau_1-y\big)}{\big(\tau_2-y\big)} \Bigg\},
\end{eqnarray}
where

\begin{equation}
         \label{eq:165}
         z(\tau,y)\equiv 2 \sqrt{a\left[v,v-(\tau+y)\right]\, \left(\tau-y\right)}.
\end{equation}
Comparing \eqref{eq:163} and \eqref{eq:164} one can see that they are identical. Indeed, if we exchange $\tau_{1} \leftrightarrow \tau_{2}$ in $[G^K_{an2}]_{22}$ and use that $f\big[\sqrt{2}(\tau_{1}-\tau_{2})\big] = - f\big[\sqrt{2}(\tau_{2}-\tau_{1})\big]$  we obtain $[G^K_{an2}]_{11}$.

Furthermore, non--diagonal components of $G^K_{an2}$ have the following form:

\begin{eqnarray}
    \label{eq:d166}
   [G^K_{an2}]_{12}(x,t;x,t) = \nonumber \\
   = -i \, \int_{0}^{+\infty} d\tau_1 \int_0^{+\infty} d\tau_2 \int_{-min(\tau_1,\tau_2)}^{min(\tau_1,\tau_2)}dy \, \bigg\{\Phi\left[v-(\tau_2 + y)\right] - \Phi\left[v-(\tau_1 + y)\right]\bigg\} \times \nonumber \\
   \times f\left[\sqrt{2}(\tau_1 - \tau_2)\right] \, \left\{\Phi\left[v - (\tau_1 + y)\right] \, J_0\left[z(\tau_1,y)\right] \, J_1\left[z(\tau_2,y)\right] \, \frac{z(\tau_2,y)}{2(\tau_2-y)} \right. - \nonumber \\
   \left. - \Phi\left[v-(\tau_2 + y)\right] \, J_0\left[z(\tau_2,y)\right] \, J_1\left[z(\tau_1,y)\right] \, \frac{z(\tau_1,y)}{2(\tau_1-y)} \right\} ,
\end{eqnarray}
and
\begin{eqnarray}
    \label{eq:d167}
   [G^K_{an2}]_{21}(x,t;x,t) = \nonumber \\
   = i \, \Phi^{2}(v) \, \int_0^{+\infty} d\tau_1 \, \int_0^{+\infty} d\tau_2 \,  \int_{-min(\tau_1,\tau_2)}^{min(\tau_1,\tau_2)} dy \, \bigg\{\Phi\left[v-(\tau_2 + y)\right] - \Phi\left[v-(\tau_1 + y)\right] \bigg\} \times \nonumber \\
   \times f\left[\sqrt{2}(\tau_1 - \tau_2)\right] \, \left\{\Phi\left[v-(\tau_1 + y)\right] \, J_0\left[z(\tau_2,y)\right] \, J_1\left[z(\tau_1,y)\right] \, \frac{2(\tau_1-y)}{z(\tau_1,y)} - \right. \nonumber \\
   - \left. \Phi\left[v-(\tau_2 + y)\right] \, J_0\left[z(\tau_1,y)\right] \, J_1\left[z(\tau_2,y)\right] \, \frac{2(\tau_2-y)}{z(\tau_2,y)}\right\}.
\end{eqnarray}
One can see that $[G^K_{an2}]_{12} = [G^K_{an2}]_{21}=0$ due to the same reason why diagonal components of $G^K_{an2}$ are equal.

Finally, we obtain that

\begin{equation}
    \label{eq:d168}
    G^K_{an2}(x,t;x,t) = [G^K_{an2}]_{11} \, \left[ {\begin{array}{cc}
   1 & 0 \\
    0 & 1 \\
  \end{array} } \right],
\end{equation}
where $[G^K_{an2}]_{11}$ is defined in eq. (\ref{eq:163}).

Consider now $G^K_{an3}(x,t;x,t)$ part of the Keldysh propagator. After straightforward manipulations, one can find that:

\begin{eqnarray}
    \label{eq:166}
    [G^K_{an3}]_{11}(x,t;x,t) = [G^K_{an3}]_{22}(x,t;x,t)= \nonumber \\
    = -\frac{\Phi\big(v\big)}{2\sqrt{2}}\int_{0}^{+\infty}d\tau\int_{0}^{+\infty}d\tau^{\prime} f\big(\sqrt{2}\tau \big)\bigg[\Phi\big(v-\tau^{\prime}\big)-\Phi\big(v-\tau^{\prime}+\tau\big)\bigg]\times \nonumber \\
    \times \Phi\big(v-\tau^{\prime}\big)\sqrt{\dfrac{\tau}{a(v+\delta,v-\tau^{\prime})}}J_{1}\bigg(2\sqrt{\tau a\big(v,v-\tau^{\prime}\big)}\bigg)- \nonumber \\
    -\dfrac{1}{\sqrt{2}}\int_{0}^{+\infty}d\tau\int_{0}^{+\infty}d\tau^{\prime}\bigg[\Phi\big(v\big)-\Phi\big(v-\tau \big)\bigg]f\big(\sqrt{2}\tau \big)\sqrt{\dfrac{a(v,v-\tau)}{\tau^{\prime}}}J_{1}\bigg(2\sqrt{\tau^{\prime} a\big(v,v-\tau \big)}\bigg).
\end{eqnarray}
To find non--diagonal components of the spinor matrix $G^K_{an3}(x,t;x,t)$ consider:

\begin{eqnarray}
    \label{eq:e167}
   \bigg(G^{R}(x,t;y,\tau_{1})G^{A}(y,\tau_{2};x,t)\bigg)^{(3)}_{12}=  \nonumber \\
    =\dfrac{i}{2}\theta\big(t-x-\tau_{1}+y\big)\delta\big(t+x-\tau_{1}-y\big)\theta\big(t-x-\tau_{2}+y\big)\theta\big(t+x-\tau_{2}-y\big)\times \nonumber \\
    \times \Phi(\tau_{2}-y)J_{0}\bigg(2\sqrt{a(\tau_{2}-y,t-x)\dfrac{\tau_{2}+y-t-x}{\sqrt{2}}}\bigg)-\nonumber \\
    -\dfrac{i}{2}\theta\big(t-x-\tau_{2}+y\big)\delta\big(t+x-\tau_{2}-y\big)\theta\big(t-x-\tau_{1}+y\big)\theta\big(t+x-\tau_{1}-y\big)\times \nonumber \\
    \times \Phi(\tau_{1}-y)J_{0}\bigg(2\sqrt{a(\tau_{1}-y,t-x)\dfrac{\tau_{1}+y-t-x}{\sqrt{2}}}\bigg) \ .
\end{eqnarray}
From \eqref{eq:e167} it follows that

\begin{equation}
    \label{eq:e168}
    \dfrac{1}{2}\bigg[\bigg(G^{R}(x,t;y,\tau_{1})G^{A}(y,\tau_{2};x,t)\bigg)^{(3)}+\bigg(G^{R}(x,t;y,\tau_{2})G^{A}(y,\tau_{1};x,t)\bigg)^{(3)}\bigg]=0.
\end{equation}
Hence,

$$[G^K_{an3}]_{12}(x,t;x,t)=0.$$
The same arguments are valid also for:
\begin{eqnarray}
    \label{eq:e169}
   \bigg(G^{R}(x,t;y,\tau_{1})G^{A}(y,\tau_{2};x,t)\bigg)^{(3)}_{21}= \nonumber \\
   =\dfrac{i}{2}\delta\big(t-x-\tau_{1}+y\big)\theta\big(t+x-\tau_{1}-y\big)\theta\big(t-x-\tau_{2}+y\big)\theta\big(t+x-\tau_{2}-y\big) \times \nonumber \\
   \times \Phi(t-x)J_{0}\bigg(2\sqrt{a(\tau_{2}-y,t-x)\dfrac{\tau_{2}+y-t-x}{\sqrt{2}}}\bigg)- \nonumber \\
   -\dfrac{i}{2}\delta\big(t-x-\tau_{2}+y\big)\theta\big(t+x-\tau_{2}-y\big)\theta\big(t-x-\tau_{1}+y\big)\theta\big(t+x-\tau_{1}-y\big) \times \nonumber \\
   \times \Phi(t-x)J_{0}\bigg(2\sqrt{a(\tau_{1}-y,t-x)\dfrac{\tau_{1}+y-t-x}{\sqrt{2}}}\bigg) \ .
\end{eqnarray}
As a result,

$$[G^K_{an3}]_{21}(x,t;x,t)=0.$$
Finally, we find that

\begin{equation}
    \label{eq:e170}
    G^K_{an3}(x,t;x,t) = [G^K_{an3}]_{11}(x,t;x,t) \, \left[ {\begin{array}{cc}
   1 & 0 \\
    0 & 1 \\
  \end{array} } \right] \ ,
\end{equation}
where $[G^{K}_{an3}]_{11}(x,t;x,t)$ is defined in eq. (\ref{eq:166}). Combining all the results of this Appendix together we obtain the eq.(\ref{GKth}).

\section{Calculation of the scalar current in the functional formalism} \label{D}

To find the leading contribution to the scalar current coming from $[G^K_{th}]_{11}$ in the limit \eqref{eq:100} we have to consider the two types of integrals in eq. \eqref{eq:62}. To estimate them it is convenient to introduce the dimensionless variable:

\begin{equation}
    \label{eqn1}
    z=\tau \lambda \ .
\end{equation}
In terms of $z$ the integrals in \eqref{eq:62} acquire the following form:

\begin{equation}
    \label{eqn2}
    I_{1}(\delta,\infty) \equiv \int_{\delta \lambda}^{+\infty} dz\dfrac{1}{z}J_{0}\bigg(2\sqrt{ \dfrac{z}{2\lambda^{2}}\int_{\lambda v-z}^{\lambda v}dy \Phi^{2}(y/\lambda)}\bigg) \
\end{equation}
and
\begin{equation}
    \label{eqn3}
    I_{2}(\delta,\infty) \equiv \int_{\delta \lambda}^{+\infty} dz\dfrac{\Phi(v-z/\lambda)}{z}J_{0}\bigg(2\sqrt{ \dfrac{z}{2\lambda^{2}}\int_{\lambda v-z}^{\lambda v}dy \Phi^{2}(y/\lambda)}\bigg),
\end{equation}
if $f(\tau)$ is taken to be as in eq. (\ref{eq:k1}).

First of all, in the limit \eqref{eq:100} we have that:

\begin{equation}
    \label{eq:103}
    \dfrac{1}{2\lambda^{2}}\int_{\lambda v-z}^{\lambda v}dz \, \Phi^{2}(z/\lambda)=\dfrac{1}{2}z \dfrac{\Phi^{2}(v)}{\lambda^{2}}\bigg[1+\sum_{n=1}a_{n}z^{n}\bigg] \ ,
\end{equation}
where
$$a_{n}=\dfrac{(-1)^{n}}{\lambda^{n} n!}\dfrac{1}{\Phi^{2}(v)}\dfrac{d^{n}}{dv^{n}}\Phi^{2}(v), \qquad \qquad \text{and} \qquad \qquad \big|a_{n}\big| < 1 \ .$$
Consider now the integral as follows:

\begin{equation}
    \label{eq:104}
    I_{1}(\delta,1)= \int_{\delta \lambda}^{1} dz\dfrac{1}{z}J_{0}\bigg(2\sqrt{ \dfrac{z}{2\lambda^{2}}\int_{\lambda v-z}^{\lambda v}dy \Phi^{2}(y/\lambda)}\bigg)=\int_{\delta\lambda}^{1} dz\dfrac{1}{z}J_{0}\bigg(A z \sqrt{1+\sum_{n=1}a_{n}z^{n}}\bigg),
\end{equation}
where for convenience we use the following notation: $A=\sqrt{2}\dfrac{|\Phi(v)|}{\lambda}$. And please note the change of the upper limit of integration in (\ref{eq:104}) with respect to (\ref{eqn2}). We expand the Bessel function $J_{0}$ in terms of $a_{i}$, using that:

\begin{equation}
    \label{eq:105}
    \partial_{i_{1}}...\partial_{i_{n}}J_{0}\bigg(Az \sqrt{1+\sum_{n=1}a_{n}z^{n}}\bigg)\bigg|_{a_{i}=0}=\dfrac{(-1)^{n}}{2^{n}}\big(Az\big)^{n}J_{n}\big(A z\big)z^{i_{1}}...z^{i_{n}} \ .
\end{equation}
Then,
\begin{eqnarray}
    \label{eq:106}
    I_{1}(\delta,1)=\int_{\delta\lambda}^{1} dz\dfrac{1}{z}J_{0}\bigg(Az\bigg)+\sum_{i_{1}=1}a_{i_{1}}\dfrac{(-1)}{2}A\int_{0}^{1}d z \  z^{i_{1}+1-1}J_{1}(Az)+ \nonumber \\
    +\sum_{i_{1},i_{2}=1}\dfrac{1}{2!}a_{i_{1}}a_{i_{2}}\dfrac{(-1)^{2}}{2^{2}}A^{2}\int_{0}^{1} dz \ z^{i_{1}+i_{2}+2-1}J_{2}(Az)+\nonumber \\
    +\sum_{i_{1},i_{2},i_{3}=1}\dfrac{1}{3!}a_{i_{1}}a_{i_{2}}a_{i_{3}}\dfrac{(-1)^{3}}{2^{3}}A^{3}\int_{0}^{1} dz \ z^{i_{1}+i_{2}+i_{3}+3-1}J_{3}(Az)+... \ .
\end{eqnarray}
For $A \gg 1$ the following approximation is true \cite{1}:

\begin{eqnarray}
    \label{eq:107}
    \int_{0}^{1} z^{\mu}J_{\nu}\big(Az\big)dz \approx -\sqrt{\dfrac{2}{\pi}}\dfrac{\cos\bigg(A-\pi(2\nu+1)/4\bigg)}{A^{3/2}}+\dfrac{1}{A^{3}}+\nonumber \\
    +\sqrt{\dfrac{2}{\pi}}\dfrac{(\nu+\mu-1)}{A^{\mu+1/2}}\cos\bigg(A-\pi(2\nu+1)/4\bigg)+\dfrac{2^{\mu}}{A^{\mu+1}}\dfrac{\Gamma \bigg(1/2(1+\mu+\nu)\bigg)}{\Gamma \bigg(1/2(1-\mu+\nu)\bigg)} \ .
\end{eqnarray}
Using \eqref{eq:107}, one can estimate that:
\begin{equation}
    \label{eq:108}
    I_{1}(\delta,1)=\int_{\delta\lambda}^{1} dz\dfrac{1}{z}J_{0}\bigg(Az\bigg) + \dfrac{1}{
    A^{3}}e^{B}-\dfrac{a_{1}}{\sqrt{2\pi A}} \cos\bigg(A-3\pi/4\bigg) + o\left(\dfrac{1}{A^{3/2}}\right) \ ,
\end{equation}
where
\begin{equation}
    \label{eq:109}
    B=\dfrac{A}{2}\sum_{i}a_{i}=-\dfrac{1}{\sqrt{2}\Phi(v)}\bigg(\dfrac{\Phi^{2}(v)}{\lambda}-\int_{v-1/\lambda}^{v}\Phi^{2}(y)dy\bigg) \ .
\end{equation}
In our case the difference $\dfrac{\Phi^{2}(v)}{\lambda}-\int_{v-1/\lambda}^{v}\Phi^{2}(y)dy$ is not of order $\Phi^{2}(v)/\lambda$, since $\Phi$ is a very slow function of its argument, so we can neglect the term $\dfrac{1}{A^{3}} e^{B}$ and obtain that:

\begin{equation}
    \label{eq:110}
    \int_{\delta \lambda}^{1} dz\dfrac{1}{z}J_{0}\bigg(Az\bigg)=\int_{A\delta\lambda}^{A} dz\dfrac{1}{z}J_{0}\bigg(z\bigg)\approx\int_{A\delta \lambda}^{\infty} dz\dfrac{1}{z}J_{0}\bigg(z\bigg)\approx -\ln\bigg(\dfrac{\delta A\lambda}{2}\bigg)-\gamma \ .
\end{equation}
One can change the upper limit of integration, since for $A \gg 1$:
$$\int_{A}^{\infty}\dfrac{dz}{z}J_{0}(z)\approx \int_{A}^{\infty}\dfrac{dz}{z\sqrt{z}}\cos(z) \approx -\dfrac{\sin{A}}{A^{3/2}}+o(\dfrac{1}{A^{5/2}}) \ . $$
Thus, we obtain that:

\begin{equation}
    \label{eq:111}
    I_{1}(\delta,1)=-\ln\bigg(\dfrac{\delta |\Phi(v)|}{2}\bigg)-\gamma+\dfrac{\Phi^{\prime}(v)}{\lambda \Phi(v)}\sqrt{\dfrac{\lambda}{4\pi |\Phi(v)|}} \cos\bigg(\sqrt{2}\dfrac{|\Phi(v)|}{\lambda}-3\pi/4\bigg)+o\bigg(\dfrac{\lambda^{3/2}}{|\Phi(v)|^{3/2}}\bigg) \ .
\end{equation}
The next step is to estimate the remaining part of the integral from eq.(\ref{eqn2}):

\begin{equation}
    \label{eq:112}
    I_{1}(1,\infty)=\int_{1}^{\infty} dz\dfrac{1}{z}J_{0}\bigg(2\sqrt{ \dfrac{z}{2\lambda^{2}}\int_{\lambda v-z}^{\lambda v}dy \Phi^{2}(y/\lambda)}\bigg) \ .
\end{equation}
We use, that $\dfrac{1}{\lambda^{2}}\int_{\lambda v-1}^{\lambda v}dy \ \Phi^{2}(y/\lambda) \gg 1$ and, hence,
$$\dfrac{1}{\lambda^{2}}\int_{\lambda v-z}^{\lambda v}dy \ \Phi^{2}(y/\lambda)=\dfrac{1}{\lambda^{2}}\int_{\lambda v-z}^{\lambda v-1}dy \ \Phi^{2}(y/\lambda)+\dfrac{1}{\lambda^{2}}\int_{\lambda v-1}^{\lambda v}dy \ \Phi^{2}(y/\lambda)=\dfrac{1}{\lambda^{2}}\int_{\lambda v-1}^{\lambda v}dy \ \Phi^{2}(y/\lambda)\bigg(1+g(\tau)\bigg) \ ,$$
where
$$g(\tau)=\dfrac{\int_{\lambda v-z}^{\lambda v-1}dy \ \Phi^{2}(y/\lambda)}{\int_{\lambda v-1}^{\lambda v}dy \ \Phi^{2}(y/\lambda)}>0 \ .$$
Using asymptotic expansion for the Bessel function $J_{0}$, we obtain that:
\begin{eqnarray}
    \label{eq:113}
    |I_{1}|\approx \left|\int_{1}^{\infty}dz\dfrac{1}{z}\dfrac{\cos\bigg(\sqrt{2z\dfrac{1}{\lambda^{2}}\int_{\lambda v-z}^{\lambda v}dy \ \Phi^{2}(y/\lambda)}\bigg)}{\bigg[2z\dfrac{1}{\lambda^{2}}\int_{\lambda v-z}^{\lambda v}dy \ \Phi^{2}(y/\lambda)\bigg]^{1/4}}\right| < \dfrac{1}{\bigg[\dfrac{1}{\lambda^{2}}\int_{\lambda v-1}^{\lambda v}dy \ \Phi^{2}(y/\lambda)\bigg]^{1/4}}\int_{1}^{\infty}dz\dfrac{1}{z}\dfrac{1}{z^{1/4}}= \nonumber \\
    =\dfrac{4}{\bigg[\dfrac{1}{\lambda^{2}}\int_{\lambda v-1}^{\lambda v}dy \ \Phi^{2}(y/\lambda)\bigg]^{1/4}} \sim \sqrt{\dfrac{\lambda}{\Phi(v)}} \ . \ \ \
\end{eqnarray}
Then, the total result for (\ref{eqn2}) is:

\begin{equation}
    \label{eq:114}
    I_{1}(\delta,\infty)=-\ln\bigg(\dfrac{\delta |\Phi(v)|}{2}\bigg)-\gamma+\dfrac{\Phi^{\prime}(v)}{\lambda \Phi(v)}\sqrt{\dfrac{\lambda}{4\pi |\Phi(v)|}} \cos\bigg(\sqrt{2}\dfrac{|\Phi(v)|}{\lambda}-3\pi/4\bigg)+o\bigg(\dfrac{\lambda^{3/2}}{|\Phi(v)|^{3/2}}\bigg) \ ,
\end{equation}
and also

\begin{equation}
    \label{eq:115}
    I_{2}(\delta,1)=\Phi(v)I_{1}(\delta,1)+o\bigg(\dfrac{\lambda^{3/2}}{|\Phi(v)|^{3/2}}\bigg),
\end{equation}
in the limit in question.

The remaining integral that we have to estimate is as follows:
\begin{eqnarray}
    \label{eq:116}
    I_{2}(1,\infty)=\int_{1}^{+\infty} dz \dfrac{\Phi(v-z/\lambda)}{z}J_{0}\bigg(\sqrt{2z\dfrac{1}{\lambda^{2}}\int_{\lambda v-z}^{\lambda v}dy \ \Phi^{2}(y/\lambda)}\bigg)\approx \nonumber \\
    \approx \int_{1}^{\infty}dz \ \dfrac{\Phi(v-z/\lambda)}{z}\dfrac{\cos\bigg(\sqrt{2z\dfrac{1}{\lambda^{2}}\int_{\lambda v-z}^{\lambda v}dy \ \Phi^{2}(y/\lambda)}\bigg)}{\bigg[2z\dfrac{1}{\lambda^{2}}\int_{\lambda v-z}^{\lambda v}dy \ \Phi^{2}(y/\lambda)\bigg]^{1/4}} \ .
\end{eqnarray}
It obeys the following conditions:

\begin{eqnarray}
    \label{eq:117}
     \left|\int_{1}^{\infty}dz \ \dfrac{\Phi(v-z/\lambda)}{z}\dfrac{\cos\bigg(\sqrt{2z\dfrac{1}{\lambda^{2}}\int_{\lambda v-z}^{\lambda v}dy \ \Phi^{2}(y/\lambda)}\bigg)}{\bigg[2z\dfrac{1}{\lambda^{2}}\int_{\lambda v-z}^{\lambda v}dy \ \Phi^{2}(y/\lambda)\bigg]^{1/4}}\right|<\nonumber \\
     <\dfrac{1}{\bigg[\dfrac{1}{\lambda^{2}}\int_{\lambda v-1}^{\lambda v}dy \ \Phi^{2}(y/\lambda)\bigg]^{1/4}}\int_{1}^{\infty} dz \ \dfrac{\Phi(v-z/\lambda)}{z}\dfrac{1}{z^{1/4}}= \nonumber \\ = \dfrac{4M}{\bigg[\dfrac{1}{\lambda^{2}}\int_{\lambda v-1}^{\lambda v}dy \ \Phi^{2}(y/\lambda)\bigg]^{1/4}} \sim \dfrac{M\sqrt{\lambda}}{\sqrt{\Phi(v)}} \sim \sqrt{\lambda \Phi(v)} \ ,
\end{eqnarray}
where
\begin{equation}
    \label{eq:118}
    M=\max_{z \in (1,\infty)}\bigg(\Phi(v-z/\lambda)\bigg) \sim \Phi(v) \ .
\end{equation}
In all, we find that the expression \eqref{eq:62} in the limit (\ref{eq:100}) can be approximated as:

\begin{equation}
    \label{eq:119}
     [G^K_{th}]_{11}(x,t;x-\delta,t) = \dfrac{i}{\pi}\Phi(v)\ln\bigg(\dfrac{\delta |\Phi(v)|}{2}\bigg)+\dfrac{i}{\pi}\Phi(v)\gamma+
\end{equation}
$$+\dfrac{i}{\pi}\dfrac{\Phi^{\prime}(v)}{\lambda \Phi(v)}\sqrt{\dfrac{\lambda|\Phi(v)|}{4\pi }} \cos\bigg(\sqrt{2}\dfrac{|\Phi(v)|}{\lambda}-3\pi/4\bigg)+o\bigg(\dfrac{\lambda^{3/2}}{|\Phi(v)|^{1/2}}\bigg) \ .$$
To find the leading contribution to the correlation function coming from $[G^K_{an2}]_{11}$ in the limit (\ref{eq:100}) we need to estimate the integral in eq.(\ref{eq:163}). One can see that if $\Phi(v)=const$ this integral vanishes. Hence, it is proportional to some derivative of the potential and contributes subleading corrections to the scalar current in the limit (\ref{eq:100}).

To find the leading contribution to the scalar current from $[G^K_{an3}]_{11}$ in the limit (\ref{eq:100}) one needs to estimate the integral
in eq. (\ref{eq:166}). In the first row of (\ref{eq:166}) we can take the integral over $\tau^{\prime}$ explicitly:

\begin{equation}
    \label{eq:K(3)1}
    \int_{0}^{+\infty}d\tau^{\prime}\sqrt{\dfrac{a(v,v-\tau)}{\tau^{\prime}}}J_{1}\bigg(2\sqrt{\tau^{\prime} a\big(v,v-\tau \big)}\bigg)=\int_{0}^{+\infty}dy \ J_{1}(y)=1 \ ,
\end{equation}
so,

\begin{eqnarray}
    \label{eq:K(3)2}
    \dfrac{1}{\sqrt{2}}\int_{\delta}^{+\infty}d\tau\int_{0}^{+\infty}d\tau^{\prime}\bigg[\Phi\big(v\big)-\Phi\big(v-\tau \big)\bigg]f\big(\sqrt{2}\tau \big)\sqrt{\dfrac{a(v,v-\tau)}{\tau^{\prime}}}J_{1}\bigg(2\sqrt{\tau^{\prime} a\big(v,v-\tau \big)}\bigg)= \nonumber \\ = -\dfrac{i}{2\pi}\int_{0}^{+\infty}d\tau\dfrac{1}{\tau+\delta} \bigg[\Phi\big(v\big)-\Phi\big(v-\tau \big)\bigg],
\end{eqnarray}
where we have used the expression for $f(\tau)$ from eq. (\ref{eq:k1}).

Thus, the first integral on the right hand side of (\ref{eq:166}) can be neglected, since we assume that $\Phi(x)$ is a very slow function of $x$, which means that the difference $\Phi(v-\tau^{\prime})-\Phi\big(v-\tau^{\prime}+\tau\big)$ is small when $\delta<\tau<1$.

The other integral in (\ref{eq:166}) can be represented in the following form:

\begin{eqnarray}
    \label{eq:K(3)4}
    -\dfrac{\Phi\big(v\big)}{2\sqrt{2}}\int_{1}^{+\infty}d\tau\int_{0}^{+\infty}d\tau^{\prime}\bigg[\Phi\big(v-\tau^{\prime}\big)\bigg]f\big(\sqrt{2}\tau \big)\Phi\big(v-\tau^{\prime}\big)\sqrt{\dfrac{\tau}{a(v,v-\tau^{\prime})}}J_{1}\bigg(2\sqrt{\tau a\big(v,v-\tau^{\prime}\big)}\bigg)= \nonumber \\ = \dfrac{i\Phi(v)}{2\pi}\int_{1}^{\infty}d\tau \int_{0}^{a}dy \ \dfrac{1}{\tau}\sqrt{\dfrac{t-\delta}{y}}J_{1}\big(2\sqrt{(\tau-\delta)\, a(v,v-\tau^{\prime})}\big)=\nonumber \\
    =\dfrac{i\Phi(v)}{2\pi}\int_{1}^{\infty}d\tau  \dfrac{1}{\tau+\delta}\bigg(1-J_{0}\big(2 \sqrt{a(v,-\infty)\tau}\big)\bigg).
\end{eqnarray}
To get this result we have made the change of variables $s=2\sqrt{(\tau-\delta) \, a(v,v-\tau^{\prime})}$ and used that $\int_{0}^{b}dx \ J_{1}(x)=1-J_{0}(b)$. Then:

\begin{eqnarray}
    \label{eq:SDEK(3)4}
    -\dfrac{\Phi\big(v\big)}{2\sqrt{2}}\int_{1}^{+\infty}d\tau\int_{0}^{+\infty}d\tau^{\prime}\bigg[\Phi\big(v-\tau^{\prime}+\tau\big)\bigg]f\big(\sqrt{2}\tau \big)\Phi\big(v-\tau^{\prime}\big)\sqrt{\dfrac{\tau}{a(v,v-\tau^{\prime})}}J_{1}\bigg(2\sqrt{\tau a\big(v,v-\tau^{\prime}\big)}\bigg)\approx \nonumber \\
    \approx -\dfrac{\Phi\big(v\big)}{2\sqrt{2}}\int_{1}^{s}d\tau\int_{0}^{s^{\prime}}d\tau^{\prime}\bigg[\Phi(v)\bigg]f\big(\sqrt{2}\tau \big)\Phi(v)\sqrt{\dfrac{\tau}{1/2\Phi^{2}(v)\tau^{\prime}}}J_{1}\bigg(2\sqrt{\tau \ 1/2\Phi^{2}(v)\tau^{\prime}}\bigg)\approx \nonumber \\
    \approx \dfrac{\Phi^{3}(v)}{\Phi(v)}\int_{1}^{s}\dfrac{d\tau}{\tau}\int_{0}^{s^{\prime}}d \tau^{\prime}\sqrt{\dfrac{\tau}{\tau^{\prime}}}J_{1}\bigg(\Phi(v)\sqrt{2\tau \ \tau^{\prime}}\bigg)\approx \nonumber \\
    \approx \dfrac{\Phi(v)}{s^{\prime}}\int_{1}^{s}\dfrac{d\tau}{\tau}\bigg[1-J_{0}\bigg(\Phi(v)\sqrt{\tau s^{\prime}}\bigg)\bigg] \approx \dfrac{\Phi(v)}{s^{\prime}}\ln{s} \ , \ \ \
\end{eqnarray}
where in \eqref{eq:SDEK(3)4} we have made the assumption that the leading contribution to the integral is coming from such values of $s$ and $s^{\prime}$ which correspond to some finite values of $\tau$ and $\tau^{\prime}$ respectively, such that when $ 1<\tau<s$ and $ 0<\tau^{\prime}<s^{\prime}$ one can use the approximation $\Phi(v-\tau^{\prime}+\tau)\Phi(v-\tau^{\prime}) \approx \Phi^{2}(v)$. So we can also neglect this term with respect to $\Phi(v)\ln{\big[\delta \Phi(v)\big]}$ from eq. (\ref{eq:119}).

In all, we find that the approximate value of the integral (\ref{eq:166}) is as follows:

\begin{eqnarray}
    \label{eq:K(3)5}
     G_{11}^{(\text{anom})K(3)}(x,t;x,t) \approx \nonumber \\
     \approx \dfrac{i\Phi(v)}{2\pi}\int_{1}^{\infty}d\tau  \dfrac{1}{\tau+\delta}\bigg(1-J_{0}\big(2\sqrt{a\tau}\big)\bigg)- \dfrac{i}{2\pi}\int_{0}^{+\infty}d\tau\dfrac{1}{\tau+\delta} \bigg[\Phi\big(v\big)-\Phi\big(v-\tau \big)\bigg] \approx \nonumber \\
   \approx \dfrac{i}{2\pi}\int_{0}^{1} d\tau \dfrac{\Phi(v-\tau)-\Phi(v)}{\tau+\delta}+\dfrac{i}{2\pi}\int_{1}^{\infty} d\tau \  \dfrac{\Phi(v-\tau)}{\tau+\delta}-\nonumber \\
   -\dfrac{i\Phi(v)}{2\pi}\int_{1}^{\infty}d\tau  \dfrac{1}{\tau+\delta}J_{0}\big(2\sqrt{a\tau}\big) \approx i\Phi(v) \, {\rm const.} + ...
\end{eqnarray}
where the ellipses include terms which are proportional to $\Phi^{\prime}(v)$ and $\dfrac{1}{\Phi(v)}$. Equation \eqref{eq:K(3)5} shows that in the limit (\ref{eq:100})
we can neglect $[G^K_{an3}]_{11}$ with respect to $G^{K}_{th}$. Combining the results of the calculations in this Appendix we obtain the eq. (\ref{eq:13000}) in the limit (\ref{eq:100}).

\section{Massive fermions and Bogoliubov transformation}\label{E}

The action of free 2D massive fermions is

\begin{equation}
    \label{eq:c1}
     S[\psi, \bar{\psi}]= \int dt \int dx \, \bar{\psi}(x,t)\bigg(i \slashed{\partial}-m\bigg)\psi(x,t).
\end{equation}
The equations of motion are as follows:

\begin{equation}
    \label{eq:c2}
  \left[ {\begin{array}{cc}
   -m & i\bigg(\partial_{t}+\partial_{x}\bigg) \\
    i\bigg(\partial_{t}-\partial_{x}\bigg) & -m \\
  \end{array} } \right]\begin{bmatrix}\psi_{1} \\ \psi_{2} \end{bmatrix}=0 \ .
\end{equation}
Their plane wave solutions are:

\begin{equation}
    \label{eq:c3}
    \psi=u(p)e^{-iw_{p}t+ipx} \qquad \text{and} \qquad \qquad\psi=v(p)e^{iw_{p}t-ipx} \ ,
\end{equation}
where
\begin{equation}
    \label{eq:c5}
    u(p)=\dfrac{1}{\sqrt{2}}\sqrt{\dfrac{w_{p}+m}{2w_{p}}}\begin{bmatrix} 1-\dfrac{p}{m+w_{p}} \\ 1+\dfrac{p}{m+w_{p}} \end{bmatrix} \ , \qquad \qquad \qquad v(p)=\dfrac{1}{\sqrt{2}}\sqrt{\dfrac{w_{p}+m}{2w_{p}}}\begin{bmatrix} \dfrac{p}{m+w_{p}}-1\\ 1+\dfrac{p}{m+w_{p}} \end{bmatrix} \ ,
\end{equation}
and $ w_{p}=\sqrt{p^{2}+m^{2}}$.
The field operator has the form:
\begin{equation}
    \label{eq:c4}
    \hat{\psi}(t,x)=\int_{-\infty}^{+\infty}\dfrac{dp}{2\pi}\bigg[\hat{a}_{p}u(p)e^{-iw_{p}t+ipx}+\hat{b}_{p}^{\dagger}v(p)e^{iw_{p}t-ipx}\bigg] \ ,
\end{equation}
and

$$\{\hat{a}_{p},\hat{a}_{q}^{\dagger}\}=2\pi \delta(p-q)  \ ,\qquad \qquad \qquad \{\hat{b}_{p},\hat{b}_{q}^{\dagger}\}=2\pi \delta(p-q) \ .$$
One can  check that
\begin{equation}
    \label{eq:c6}
    \{\hat{\psi}(t,x),\hat{\psi}^{\dagger}(t,y)\} = \delta(x-y)\hat{I} \ ,
\end{equation}
and
\begin{equation}
    \label{eq:c7}
    :\hat{H}:=\int_{-\infty}^{+\infty}\dfrac{dp}{2\pi}\sqrt{p^{2}+m^{2}}\bigg(\hat{a}_{p}^{\dagger}\hat{a}_{p}+\hat{b}_{p}^{\dagger}\hat{b}_{p}\bigg) \ .
\end{equation}
But we can choose another basis of modes instead of the plane waves \eqref{eq:c3}:

\begin{equation}
    \label{eq:c8}
    \psi=\tilde{u}(q)e^{-iqu-im^{2}v/2q} \qquad \text{and} \qquad  \psi=\tilde{v}(q)e^{iqu+im^{2}v/2q},
\end{equation}
where
\begin{equation}
    \label{eq:c9}
    \tilde{u}(q)=\begin{bmatrix}1 \\ \dfrac{m}{\sqrt{2}q} \end{bmatrix} \ , \qquad \qquad  \tilde{v}(q)=\begin{bmatrix}1 \\ -\dfrac{m}{\sqrt{2}q} \end{bmatrix}.
\end{equation}
As well as (\ref{eq:c3}) these functions also solve eq. \eqref{eq:c2}.

Let us find now the Bogoliubov canonical transformation between the modes \eqref{eq:c3} and \eqref{eq:c8}:
\begin{equation}
    \label{eq:c10}
    \begin{cases}
u(p)e^{-iw_{p}t+ipx}=\int_{0}^{+\infty} dq \ F(q,p)\tilde{u}(q)e^{-iqu-im^{2}v/2q} \\
v(p)e^{iw_{p}t-ipx}=   \ \int_{0}^{+\infty} dq \ G(q,p)\tilde{v}(q)e^{iqu+im^{2}v/2q}  \ .
\end{cases}
\end{equation}
It is straightforward to see that:

\begin{equation}
 \label{eq:c11}
\begin{cases}
F(q,p)=\sqrt{\dfrac{m^{2}+2q^{2}}{4q^{2}}}\delta \bigg(p-\dfrac{m^{2}-2q^{2}}{2\sqrt{2}q}\bigg)\\
G(q,p)=-\sqrt{\dfrac{m^{2}+2q^{2}}{4q^{2}}}\delta \bigg(p-\dfrac{m^{2}-2q^{2}}{2\sqrt{2}q}\bigg) \ .
\end{cases}
\end{equation}
From \eqref{eq:c11} one can observe that while $p$ takes values in the interval $(-\infty,+\infty)$, the Fourier parameter $q$ in (\ref{eq:c8}) is ranging in the interval $(0,+\infty)$. The peculiarity of the modes (\ref{eq:c8}) at $q=0$ and how to deal with it is discussed in the subsection 3.1.

Then, using \eqref{eq:c10} we can rewrite \eqref{eq:c4} as follows:
\begin{equation}
    \label{eq:c12}
    \hat{\psi}(t,x)=\int_{-\infty}^{+\infty}\dfrac{dp}{2\pi}\bigg[\hat{a}_{p}u(p)e^{-iw_{p}t+ipx}+\hat{b}_{p}^{\dagger}v(p)e^{iw_{p}t-ipx}\bigg]=
\end{equation}
$$=\int_{-\infty}^{+\infty}\dfrac{dp}{2\pi}\bigg[\hat{a}_{p}\int_{0}^{+\infty} dq \ F(q,p)\tilde{u}(q)e^{-iqu-im^{2}v/2q}+\hat{b}_{p}^{\dagger}\int_{0}^{+\infty} dq \ G(q,p)\tilde{v}(q)e^{iqu+im^{2}v/2q} \bigg]=$$
$$=\int_{0}^{+\infty}\dfrac{dq}{2\pi}\dfrac{1}{\sqrt[4]{2}}\bigg[\hat{\tilde{a}}_{q}\begin{bmatrix}1 \\ \dfrac{m}{\sqrt{2}q} \end{bmatrix}e^{-iqu-im^{2}v/2q}+\hat{\tilde{b}}_{p}^{\dagger}\begin{bmatrix}1 \\ -\dfrac{m}{\sqrt{2}q} \end{bmatrix}e^{iqu+im^{2}v/2q}\bigg],$$
where
 \begin{equation}
    \label{eq:c13}
    \begin{cases}
\hat{\tilde{a}}_{q}=\sqrt[4]{2}\sqrt{\dfrac{m^{2}+2q^{2}}{4q^{2}}}\hat{a}\bigg(\dfrac{m^{2}-q^{2}}{2\sqrt{2}q}\bigg) \\
\hat{\tilde{b}}_{q}=-\sqrt[4]{2}\sqrt{\dfrac{m^{2}+2q^{2}}{4q^{2}}}\hat{b}\bigg(\dfrac{m^{2}-q^{2}}{2\sqrt{2}q}\bigg)
\end{cases} \qquad  \text{and} \qquad \begin{cases}
\{\hat{\tilde{a}}_{p},\hat{\tilde{a}}_{q}^{\dagger}\}=2\pi \delta(p-q) \\
\{\hat{\tilde{b}}_{p},\hat{\tilde{b}}_{q}^{\dagger}\}=2\pi \delta(p-q) \ .
\end{cases}
\end{equation}
The anticommutation relations are as follows:
\begin{equation}
    \label{eq:c15}
    \{\hat{\psi}(t,x),\hat{\psi}^{\dagger}(t,y)\}=\int_{0}^{+\infty}\dfrac{dp}{\pi} \dfrac{1}{p^{2}} \cos{\bigg(\dfrac{(x-y)m^{2}}{4}p-\dfrac{(x-y)}{p}\bigg)}\left[ {\begin{array}{cc}
   1 & 0 \\
   0 & 1\\
  \end{array} } \right] \ .
\end{equation}
The integral here can be calculated as \cite{1}:

\begin{eqnarray}
    \label{eq:c16}
    I \equiv \int_{0}^{+\infty}\dfrac{dp}{\pi} \dfrac{1}{p^{2}} \cos{\bigg(\dfrac{(x-y)m^{2}}{4}p-\dfrac{(x-y)}{p}\bigg)}=\nonumber \\
    =\lim_{\epsilon \to 0}\int_{0}^{+\infty}\dfrac{dp}{\pi} \dfrac{1}{p^{2}+\epsilon^{2}} \cos{\bigg(\dfrac{(x-y)m^{2}}{4}p-\dfrac{(x-y)}{p}\bigg)}= \nonumber \\
    =\lim_{\epsilon \to 0}\dfrac{1}{2\epsilon} \exp{\bigg(-\dfrac{|x-y|m^{2}}{4}\epsilon-\dfrac{|x-y|}{\epsilon}\bigg)}=\lim_{\epsilon \to 0}\dfrac{1}{2\epsilon} \exp{\bigg(-\dfrac{|x-y|}{\epsilon}\bigg)}=\delta(x-y).
\end{eqnarray}
Then we obtain the canonical anticommutation relations:
\begin{equation}
    \label{eq:c19}
    \{\hat{\psi}(t,x),\hat{\psi}^{\dagger}(t,y)\}=\delta(x-y) \left[ {\begin{array}{cc}
   1 & 0 \\
   0 & 1\\
  \end{array} } \right] \ .
\end{equation}
For completeness, the free Hamiltonian in the new modes is as follows:

\begin{equation}
    \label{eq:c14}
    :\hat{H}:=\int_{0}^{+\infty}\dfrac{dq}{2\pi} \ \dfrac{2q^{2}+m^{2}}{2q}\bigg[\hat{\tilde{a}}_{q}^{\dagger}\hat{\tilde{a}}_{q} + \hat{\tilde{b}}_{q}^{\dagger}\hat{\tilde{b}}_{q}\bigg].
\end{equation}

\section{Calculation of the expectation value of the stress energy tensor in the operator formalism} \label{F}

In the light cone coordinates the components of the stress energy tensor have the following form:
\begin{equation}
    \label{e:3}
    \begin{cases}
    T_{vv}=\dfrac{1}{2}\big(T_{00}+T_{11}-2T_{01}\big)\\
    T_{uu}=\dfrac{1}{2}\big(T_{00}+T_{11}+2T_{01}\big)\\
    T_{uv}=T_{vu}=\dfrac{1}{2}\big(T_{00}-T_{11}\big) \ .
    \end{cases}
\end{equation}
Since the expectation value is divergent we use the point splitting regularization, i.e. after taking  the derivatives in the correlation function (\ref{eq:SETF2}) we put

\begin{equation}
    \label{e:20}
    \begin{cases}
    t^{\prime}=t-i\delta^{0} \\
    x^{\prime}=x-i\delta^{1}
    \end{cases} \qquad \longrightarrow  \qquad
    \begin{cases}
    u^{\prime}=u-i\delta^{u} \\
    v^{\prime}=v-i\delta^{v} \ ,
    \end{cases} \quad {\rm where} \quad Re (\delta^{u}), Re (\delta^{v}) > 0.
\end{equation}
Here $\delta^{u}$ and $\delta^{v}$ can be represented as
\begin{equation}
    \label{e:25}
    \begin{cases}
    \delta^{u}=\dfrac{1}{\sqrt{2}}\big(\delta^{0}+\delta^{1}\big) \\
    \delta^{v}=\dfrac{1}{\sqrt{2}}\big(\delta^{0}-\delta^{1}\big)
    \end{cases}, \quad {\rm and} \quad \delta^{2}=\delta_{\mu}\delta^{\mu}=2\delta^{v}\delta^{u}=\delta^{0}\delta^{0}-\delta^{1}\delta^{1} \ .
\end{equation}
Writing the expectation value of \eqref{eq:SET8} in components, we get
\begin{equation}
    \label{e:32}
    \begin{cases}
        \langle T^{01}\rangle = \dfrac{i}{4}\bigg[\big(\sqrt{2}\partial_{v}-\sqrt{2}\partial_{v^{\prime}}\big)\langle \psi_{2}^{\dagger}(u^{\prime},v^{\prime})\psi_{2}(u,v) \rangle -\big(\sqrt{2}\partial_{u}-\sqrt{2}\partial_{u^{\prime}}\big)\langle \psi_{1}^{\dagger}(u^{\prime},v^{\prime})\psi_{1}(u,v) \rangle \bigg]\\ \\
        \langle T_{00} \rangle+\langle T_{11} \rangle =\dfrac{i}{2}\bigg[\big(\sqrt{2}\partial_{v}-\sqrt{2}\partial_{v^{\prime}}\big)\langle \psi_{2}^{\dagger}(u^{\prime},v^{\prime})\psi_{2}(u,v)\rangle+\big(\sqrt{2}\partial_{u}-\sqrt{2}\partial_{u^{\prime}}\big)\langle \psi_{1}^{\dagger}(u^{\prime},v^{\prime})\psi_{1}(u,v)\rangle \bigg]\\ \\
        \langle T_{00} \rangle-\langle T_{11} \rangle =\dfrac{i}{2}\bigg[\big(\sqrt{2}\partial_{u}-\sqrt{2}\partial_{u^{\prime}}\big)\langle \psi_{2}^{\dagger}(u^{\prime},v^{\prime})\psi_{2}(u,v)\rangle+\big(\sqrt{2}\partial_{v}-\sqrt{2}\partial_{v^{\prime}}\big)\langle \psi_{1}^{\dagger}(u^{\prime},v^{\prime})\psi_{1}(u,v)\rangle \bigg] \ .
         \end{cases}
\end{equation}
We calculate each term in \eqref{e:32} separately.
First of all, using regularization \eqref{e:20}, we obtain that:

\begin{eqnarray}
    \label{e:21}
    a(v,v-i\delta^{v})=\dfrac{1}{2}\int_{v-i\delta^{v}}^{v}dy\Phi^{2}(y)=\nonumber \\
    =\dfrac{1}{2}\Phi^{2}(v)i\delta^{v}-\Phi(v)\Phi^{\prime}(v)\dfrac{(i\delta^{v})^{2}}{2}+\dfrac{\Phi^{\prime \ 2}(v)+\Phi(v)\Phi^{\prime \prime}(v)}{3!}(i\delta^{v})^{3}+...= \nonumber \\
    =i\delta^{v} b(v,\delta^{v}) \equiv i\delta^{v} \dfrac{\Phi^{2}(v)}{2}\bigg[1-i\dfrac{\Phi^{\prime}(v)}{\Phi(v)}\delta^{v}-\bigg(\dfrac{\Phi^{\prime 2}(v)}{\Phi^{2}(v)}+\dfrac{\Phi^{\prime \prime}(v)}{\Phi(v)}\bigg)\dfrac{1}{3}\delta^{v 2}+...\bigg].
\end{eqnarray}
Note that $b(v,0)=\dfrac{1}{2}\Phi^{2}(v)$.
Then, using the modes \eqref{eq:134}, one can find that:
\begin{equation}
    \label{eq:SO3}
    \langle 0| \psi_{1}^{\dagger}(u^{\prime},v^{\prime})\psi_{1}(u,v) | 0\rangle=\int_{0}^{\infty}\dfrac{dq}{2\pi}\dfrac{1}{\sqrt{2}}e^{iq(u-u^{\prime})+ia(v,v^{\prime})/q} \ ,
\end{equation}
and
\begin{equation}
    \label{eq:SO4}
    \langle 0| \psi_{2}^{\dagger}(u^{\prime},v^{\prime})\psi_{2}(u,v) | 0\rangle=\int_{0}^{\infty}\dfrac{dq}{2\pi}\dfrac{1}{\sqrt{2}}\dfrac{\Phi(v)\Phi(v^{\prime})}{2q^{2}}e^{iq(u-u^{\prime})+ia(v,v^{\prime})/q} \ .
\end{equation}
After taking the derivatives and using \eqref{e:20} and \eqref{e:21} we obtain that:
\begin{eqnarray}
    \label{eq:SO6}
    \big(\sqrt{2}\partial_{u}-\sqrt{2}\partial_{u^{\prime}}\big)\langle 0| \psi_{1}^{\dagger}(u^{\prime},v^{\prime})\psi_{1}(u,v) | 0\rangle=\nonumber \\
    =2i\int_{0}^{\infty}\dfrac{dp}{2\pi}\ p \ e^{-p\delta^{u}-\delta^{v} b(v,\delta^{v})/p}=\dfrac{i}{\pi}\bigg(\dfrac{1}{\delta^{u 2}}-\dfrac{\delta^{v} b(v,\delta^{v})}{\delta^{u}}\bigg)=\nonumber \\ =\dfrac{i}{\pi}\bigg(\dfrac{1}{\delta^{u 2}}-\dfrac{\delta^{v} \Phi^{2}(v)}{2\delta^{u}}\bigg)
    =\dfrac{2i}{\pi}\bigg(\dfrac{2}{\delta^{2}}-\dfrac{\Phi^{2}(v)}{2}\bigg)\dfrac{\delta^{v}\delta^{v}}{\delta^{2}} \ ,
\end{eqnarray}
and
\begin{eqnarray}
    \label{e:11}
    \big(\sqrt{2}\partial_{v}-\sqrt{2}\partial_{v^{\prime}}\big)\langle 0| \psi_{1}^{\dagger}(u^{\prime},v^{\prime})\psi_{1}(u,v)|0\rangle=\nonumber \\
    =i\bigg(\dfrac{\Phi^{2}(v)}{2}+\dfrac{\Phi^{2}(v-i\delta^{v})}{2}\bigg)\int_{0}^{\infty}\dfrac{dp}{2\pi}p \ e^{-p\delta^{u}-\delta^{v} b(v,\delta^{v})/p}=\nonumber \\ =\dfrac{i}{\pi}\bigg(\dfrac{\Phi^{2}(v)}{2}+\dfrac{\Phi^{2}(v-i\delta^{v})}{2}\bigg)K_{0}\bigg(\dfrac{|\Phi(v)|\sqrt{2\delta^{v}\delta^{u}}}{2}\bigg)\approx -\dfrac{i\Phi^{2}(v)}{\pi}\bigg(\ln{\bigg[\dfrac{\Phi(v)\delta}{2}\bigg]}+\gamma\bigg)  \ ,
\end{eqnarray}
and
\begin{eqnarray}
    \label{eq:SO5}
    \big[\sqrt{2}\partial_{v}-\sqrt{2}\partial_{v^{\prime}}\big]\langle 0| \psi_{2}^{\dagger}(u^{\prime},v^{\prime})\psi_{2}(u,v) | 0\rangle=\nonumber \\
    =\dfrac{1}{2}\bigg(\Phi^{\prime}(v)\Phi(v-i\delta^{v})-\Phi(v)\Phi^{\prime}(v-i\delta^{v})\bigg)\int_{0}^{\infty}\dfrac{dp}{2\pi}\dfrac{1}{p^{2}}\ e^{-p\delta^{u}-\delta^{v} b(v,\delta^{v})/p}+ \nonumber \\
    +\dfrac{i\Phi(v)\Phi(v-i\delta^{v})}{4}\bigg(\Phi^{2}(v)+\Phi^{2}(v-i\delta^{v})\bigg)\int_{0}^{\infty}\dfrac{dp}{2\pi}\dfrac{1}{p^{3}}e^{-p\delta^{u}-\delta^{v} b(v,\delta^{v})/p} \ . \
\end{eqnarray}
In the first term in \eqref{eq:SO5} we make the change $p=\dfrac{1}{q}$ of the integration variable and in the second term in \eqref{eq:SO5} the change is as follows: $p=\dfrac{b(v,\delta^{v})}{q}$. Then we obtain that:
\begin{eqnarray}
    \label{eq:SO7}
    \big[\sqrt{2}\partial_{v}-\sqrt{2}\partial_{v^{\prime}}\big]\langle 0| \psi_{2}^{\dagger}(u^{\prime},v^{\prime})\psi_{2}(u,v) | 0\rangle=\nonumber \\
    =\dfrac{1}{2}\bigg(\Phi^{\prime}(v)\Phi(v-i\delta^{v})-\Phi(v)\Phi^{\prime}(v-i\delta^{v})\bigg)\int_{0}^{\infty}\dfrac{dq}{2\pi}\ e^{-\delta^{v} b(v,\delta^{v})q-\delta^{u}/q}+ \nonumber \\
    +\dfrac{i\Phi(v)\Phi(v-i\delta^{v})}{4b^{2}(v,\delta^{v})}\bigg(\Phi^{2}(v)+\Phi^{2}(v-i\delta^{v})\bigg)\int_{0}^{\infty}\dfrac{dq}{2\pi}\ q \ e^{-q\delta^{v}-\delta^{u} b(v,\delta^{v})/q} \ . \ \ \
\end{eqnarray}
We handle each term in \eqref{eq:SO7} separately.
The first term is equal to:

\begin{eqnarray}
    \label{eq:SO13}
\dfrac{1}{2}\bigg(\Phi^{\prime}(v)\Phi(v-i\delta^{v})-\Phi(v)\Phi^{\prime}(v-i\delta^{v})\bigg)\int_{0}^{\infty}\dfrac{dq}{2\pi}\ e^{-\delta^{v} b(v,\delta^{v})q-\delta^{u}/q}= \nonumber \\
=\dfrac{1}{4\pi}\bigg(\Phi^{\prime}(v)\big(\Phi(v)-i\delta^{v} \Phi^{\prime}(v)+...\big)-\Phi(v)\big(\Phi^{\prime}(v)-i\delta^{v} \Phi^{\prime \prime}(v)+...\big)\bigg)\dfrac{1}{\delta^{v} b(v,\delta^{v})}= \nonumber \\ =\dfrac{i\delta^{v}}{4\pi}\Phi^{2}(v)\bigg(\dfrac{\Phi^{\prime \prime}(v)}{\Phi(v)}-\dfrac{\Phi^{\prime 2}(v)}{\Phi^{2}(v)}\bigg)\dfrac{1}{\delta^{v}  b(v,0)}=\dfrac{i}{2\pi}\bigg(\dfrac{\Phi^{\prime \prime}(v)}{\Phi(v)}-\dfrac{\Phi^{\prime 2}(v)}{\Phi^{2}(v)}\bigg)+O(\delta^{v}).
\end{eqnarray}
At the same time the second term is:

\begin{eqnarray}
    \label{eq:SO14}
\dfrac{i\Phi(v)\Phi(v-i\delta^{v})}{4b^{2}(v,\delta^{v})}\bigg(\Phi^{2}(v)+\Phi^{2}(v-i\delta^{v})\bigg)\int_{0}^{\infty}\dfrac{dq}{2\pi}\ q \ e^{-q\delta^{v}-\delta^{u} b(v,\delta^{v})/q}= \nonumber \\ =\dfrac{i\Phi(v)\Phi(v-i\delta^{v})}{8\pi b^{2}(v,\delta^{v})}\bigg(\Phi^{2}(v)+\Phi^{2}(v-i\delta^{v})\bigg)\bigg(\dfrac{1}{\delta^{v2}}-\dfrac{\delta^{u}\Phi^{2}(v)}{2\delta^{v}}\bigg).
\end{eqnarray}
To proceed further we also need to expand:

\begin{eqnarray}
    \label{eq:SO15}
    \Phi(v-i\delta^{v})\bigg(\Phi^{2}(v)+\Phi^{2}(v-i\delta^{v})\bigg)= \nonumber \\ =\bigg(\Phi(v)-i\delta^{v} \Phi^{\prime}(v)-\dfrac{1}{2}\Phi^{\prime \prime}(v) \delta^{v2}+...\bigg)\bigg(\Phi^{2}(v)+\bigg[\Phi(v)-i\delta^{v}\Phi^{\prime}(v)-\dfrac{1}{2}\Phi^{\prime \prime}(v) \delta^{v2}+...\bigg]^{2}\bigg)= \nonumber \\
    =\bigg(\Phi(v)-i\delta^{v} \Phi^{\prime}(v)-\dfrac{1}{2}\Phi^{\prime \prime}(v) \delta^{v2}+...\bigg)\times \nonumber \\
    \times \bigg(2\Phi^{2}(v)-2i\delta^{v} \Phi(v)\Phi^{\prime}(v)-\Phi^{\prime \ 2}(v)\delta^{v2}-\Phi(v)\Phi^{ \prime \prime}(v)\delta^{v2}+...\bigg)
    =\nonumber \\
    =2\Phi^{3}(v)-4i\delta^{v} \Phi^{3}(v)\dfrac{\Phi^{\prime}(v)}{\Phi(v)}-\delta^{v2}\Phi^{3}(v)\bigg(3\dfrac{\Phi^{\prime 2}(v)}{\Phi^{2}(v)}+2\dfrac{\Phi^{\prime \prime}(v)}{\Phi(v)}\bigg)+O(\delta^{v}) \ ,
\end{eqnarray}
and
\begin{eqnarray}
    \label{eq:SO16}
    \dfrac{1}{b^{2}(v,\delta^{v})}=\dfrac{4}{\Phi^{4}(v)}\dfrac{1}{\bigg[1-i\dfrac{\Phi^{\prime}(v)}{\Phi(v)}\delta^{v}-\bigg(\dfrac{\Phi^{\prime 2}(v)}{\Phi^{2}(v)}+\dfrac{\Phi^{\prime \prime}(v)}{\Phi(v)}\bigg)\dfrac{1}{3}\delta^{v2}+...\bigg]^{2}}= \nonumber \\
    =\dfrac{4}{\Phi^{4}(v)}\bigg[1+2i\delta^{v}\dfrac{\Phi^{\prime}(v)}{\Phi(v)}+\delta^{v2}\bigg(\dfrac{5}{3}\dfrac{\Phi^{\prime 2}(v)}{\Phi^{2}(v)}+\dfrac{2}{3}\dfrac{\Phi^{\prime \prime}(v)}{\Phi(v)}\bigg)-4\delta^{v2}\dfrac{\Phi^{\prime 2}(v)}{\Phi^{2}(v)}+O(\delta^{v})\bigg].
\end{eqnarray}
Using that
\begin{eqnarray}
    \label{eq:SO17}
    \dfrac{i\Phi(v)\Phi(v-i\delta^{v})}{8\pi b^{2}(v,\delta^{v})}\bigg(\Phi^{2}(v)+\Phi^{2}(v-i\delta^{v})\bigg)=\nonumber \\
    =\dfrac{i}{2\pi}\bigg[2-4i\delta^{v} \dfrac{\Phi^{\prime}(v)}{\Phi(v)}-\delta^{v2}\bigg(3\dfrac{\Phi^{\prime 2}(v)}{\Phi^{2}(v)}+2\dfrac{\Phi^{\prime \prime}(v)}{\Phi(v)}\bigg)\bigg]\times \nonumber \\
    \times \bigg[1+2i\delta\dfrac{\Phi^{\prime}(v)}{\Phi(v)}+\delta^{v2}\bigg(\dfrac{5}{3}\dfrac{\Phi^{\prime 2}(v)}{\Phi^{2}(v)}+\dfrac{2}{3}\dfrac{\Phi^{\prime \prime}(v)}{\Phi(v)}\bigg)-4\delta^{v2}\dfrac{\Phi^{\prime 2}(v)}{\Phi^{2}(v)}\bigg]= \nonumber \\
    =\dfrac{i}{\pi}+\dfrac{i}{6\pi}\delta^{v2}\bigg(\dfrac{\Phi^{\prime 2}(v)}{\Phi^{2}(v)}-2\dfrac{\Phi^{\prime \prime}(v)}{\Phi(v)}\bigg)+O(\delta^{v}) \ , \ \
\end{eqnarray}
and adding equations \eqref{eq:SO13} and \eqref{eq:SO17}, we obtain the following result:
\begin{eqnarray}
    \label{eq:SO18}
    \big[\sqrt{2}\partial_{v}-\sqrt{2}\partial_{v^{\prime}}\big]\langle 0| \psi_{2}^{\dagger}(u^{\prime},v^{\prime})\psi_{2}(u,v) | 0\rangle = \nonumber \\
    =\dfrac{i}{\pi}\bigg(\dfrac{1}{\delta^{v2}}-\dfrac{\delta^{u}\Phi^{2}(v)}{2\delta^{v}}\bigg)+\dfrac{i}{6\pi}\bigg(\dfrac{\Phi^{\prime 2}(v)}{\Phi^{2}(v)}-2\dfrac{\Phi^{\prime \prime}(v)}{\Phi(v)}\bigg)+\dfrac{i}{2\pi}\bigg(\dfrac{\Phi^{\prime \prime}(v)}{\Phi(v)}-\dfrac{\Phi^{\prime 2}(v)}{\Phi^{2}(v)}\bigg)+O(\delta)=\nonumber \\
    =\dfrac{i}{\pi}\bigg(\dfrac{1}{\delta^{v2}}-\dfrac{\delta^{u}\Phi^{2}(v)}{2\delta^{v}}\bigg)+\dfrac{i}{3\pi}\bigg(\dfrac{1}{2}\dfrac{\Phi^{\prime \prime}(v)}{\Phi(v)}-\dfrac{\Phi^{\prime 2}(v)}{\Phi^{2}(v)}\bigg)+O(\delta)= \nonumber \\
    =\dfrac{2i}{\pi}\bigg(\dfrac{2}{\delta^{2}}-\dfrac{\Phi^{2}(v)}{2}\bigg)\dfrac{\delta^{u}\delta^{u}}{\delta^{2}}+\dfrac{i}{3\pi}\bigg(\dfrac{1}{2}\dfrac{\Phi^{\prime \prime}(v)}{\Phi(v)}-\dfrac{\Phi^{\prime 2}(v)}{\Phi^{2}(v)}\bigg)+O(\delta^{v}) \ .
\end{eqnarray}
Finally, the last term which appears in the calculation of the expectation value of the stress  energy tensor \eqref{e:32} has the form as follows:
\begin{eqnarray}
    \label{e:10}
    \big(\sqrt{2}\partial_{u}-\sqrt{2}\partial_{u^{\prime}}\big)\langle 0| \psi_{2}^{\dagger}(u^{\prime},v^{\prime})\psi_{2}(u,v)| 0\rangle = \dfrac{\Phi(v)\Phi(v^{\prime})}{2}2i\int_{0}^{\infty}\dfrac{dq}{2\pi}\dfrac{1}{q}e^{-q\delta-\delta b(v,\delta)/q}=\nonumber \\
    =\dfrac{i\Phi(v-i\delta^{v})\Phi(v)}{\pi}K_{0}\bigg(\dfrac{|\Phi(v)|\sqrt{2\delta^{v}\delta^{u}}}{2}\bigg)\approx -\dfrac{i\Phi^{2}(v)}{\pi}\bigg(\ln{\bigg[\dfrac{\Phi(v)\delta}{2}\bigg]}+\gamma\bigg).
\end{eqnarray}
In \eqref{eq:SO6}, \eqref{e:11}, \eqref{eq:SO5} and \eqref{eq:SO7} we have used the following properties of the MacDonald functions \cite{12}:

\begin{equation}
    \label{eq:SO9}
    I_{n}(\alpha, \beta)=\int_{0}^{\infty}dp \  p ^{n-1} e^{-\alpha p-\beta/p}=2\bigg(\dfrac{\beta}{\alpha}\bigg)^{n/2}K_{n}\bigg(2\sqrt{\alpha \beta}\bigg), \qquad Re (\alpha), Re(\beta)>0
\end{equation}
with $n=0,1,2$. Since we take the limit $\delta^{v},\delta^{u} \rightarrow 0$, i.e in our case $\alpha \rightarrow 0$ and $\beta \rightarrow 0$ in the last equation, we need the following assymptotics of the MacDonald functions:

\begin{equation}
    \label{eq:SO10}
    K_{n}(z) \approx \begin{cases}
    -\ln{\dfrac{z}{2}}-\gamma, \ \ \ \ n=0 \\
    \ \ \ \ \dfrac{1}{z} , \ \ \ \ \ \ \ \ \ \ \ n=1 \\
    \dfrac{2}{z^{2}}-\dfrac{1}{2}, \ \ \ \ \ \ \ \  n=2
    \end{cases}, \quad \text{as} \quad z \rightarrow 0
\end{equation}
Then, the leading term of each integral behaves as

\begin{equation}
\label{eq:SO11}
    I_{n}(\alpha, \delta) \approx
    \begin{cases}
    -2\bigg(\ln{\sqrt{\alpha \beta}}+\gamma\bigg), \ \ \ \ n=0 \\
    \dfrac{1}{\alpha},  \ \ \ \ \ \ \ \ \ \ \ \ \ \ \ \ \ \ \ \ \ \ \ \ \ n=1 \\ \\
    \dfrac{1}{\alpha^{2}}-\dfrac{\beta}{\alpha},  \ \ \ \ \ \ \ \ \ \ \ \ \ \ \ \ \ n=2
    \end{cases}.
\end{equation}
Also, one has to keep in mind that the results obtained above are valid only if

$$ Re \, \delta^{v} b(v,\delta^{v}) = Re \,  \bigg(\delta^{v} \bigg[1-i\dfrac{\Phi^{\prime}(v)}{\Phi(v)}\delta^{v}-\bigg(\dfrac{\Phi^{\prime 2}(v)}{\Phi^{2}(v)}+\dfrac{\Phi^{\prime \prime}(v)}{\Phi(v)}\bigg)\dfrac{1}{3}\delta^{v 2}+...\bigg]\bigg)> 0 \ .$$
Particularly, this condition is satisfied when
$$\label{qe:100}
     \dfrac{1}{\lambda^{k} \Phi^{2}(v)} \dfrac{d^{k}}{dv^{k}}\Phi^{2}(v) \ll 1 \ , \qquad \text{with} \qquad k=1,2,... \ , $$
i.e. in the same limit as (\ref{eq:100}). And finally, combining together \eqref{eq:SO6}, \eqref{e:11}, \eqref{eq:SO18}, \eqref{e:10} and \eqref{e:32} we find that:

\begin{equation}
    \label{e:40}
    \begin{cases}
     \langle T^{01}(t,x)\rangle=-\dfrac{1}{\pi}\bigg(\dfrac{2}{\delta^{2}}-\dfrac{\Phi^{2}(v)}{2}\bigg)\dfrac{\delta^{0}\delta^{1}}{\delta^{2}}-\dfrac{1}{48\pi}\{\bar{a}(v),v\} \\
    \langle T^{00} \rangle+\langle T^{11} \rangle = -\dfrac{1}{\pi}\bigg(\dfrac{2}{\delta^{2}}-\dfrac{\Phi^{2}(v)}{2}\bigg)\dfrac{\delta^{0}\delta^{0}+\delta^{1}\delta^{1}}{\delta^{2}}-\dfrac{1}{24\pi}\{\bar{a}(v),v\}\\
    \langle T^{00} \rangle-\langle T^{11} \rangle =\dfrac{\Phi^{2}(v)}{\pi}\bigg(\ln{\bigg[\dfrac{\Phi(v)\delta}{2}\bigg]}+\gamma\bigg) \ .
    \end{cases}
\end{equation}
Using that $\delta^{0}\delta^{0}+\delta^{1}\delta^{1}=2\delta^{0}\delta^{0}-\delta^{2}=2\delta^{1}\delta^{1}+\delta^{2}$ we can write \eqref{e:40} as:
\begin{equation}
    \label{eq:SO19}
   \langle T^{\mu \nu} \rangle=-\dfrac{1}{\pi}\bigg(\dfrac{2}{\delta^{2}}-\dfrac{\Phi^{2}(v)}{2}\bigg)\bigg(\dfrac{\delta^{\mu}\delta^{\nu}}{\delta^{2}}-\dfrac{1}{2}\eta^{\mu \nu}\bigg)+\dfrac{\Phi^{2}(v)}{2\pi}\bigg(\ln{\bigg[\dfrac{\Phi(v)\delta}{2}\bigg]}+\gamma\bigg)\eta^{\mu \nu}-\dfrac{1}{48\pi}\{\bar{a}(v),v\}
   \left[ \begin{array}{cc}
   1& 1 \\
    1 & 1 \\
  \end{array}  \right] \ .
\end{equation}
where

$$
\bar{a}(v)=\dfrac{1}{m^{2}}\int_{}^{v}dy\Phi^{2}(y),
$$
and $\{f(z),z\}$ is the Schwarzian derivative:
\begin{equation}
    \label{eq:20}
    \{f(z),z\}=\dfrac{f^{\prime \prime \prime}(z)}{f^{\prime}(z)}-\dfrac{3}{2}\bigg(\dfrac{f^{\prime \prime}(z)}{f^{\prime}(z)}\bigg)^{2} \ .
\end{equation}
Recall that above we have taken complex regularizion parameters $\delta$'s, which can be represented as

\begin{equation}
    \label{e:34}
    \begin{cases}
    \delta^{u}=\epsilon^{u}+it^{u}\\
    \delta^{v}=\epsilon^{v}+it^{v}
    \end{cases}, \qquad \qquad
    \begin{cases}
    \epsilon^{\mu} \ll 1 \\
    t^{\mu} \ll 1 \ .
    \end{cases}
\end{equation}
where non-zero $\epsilon^{\mu}$ guarantees that all integrals which appear in the above expressions are convergent in regular sense (rather than on the space of generalized functions) and $t^{\mu}$ is a real point splitting parameter. The result \eqref{eq:SO19} was obtained when $\epsilon^{u},\epsilon^{v}>0$, but now one can safely put $\epsilon^{\mu}=0$. Then the answer for the expectation value is as follows:

\begin{equation}
    \label{e:35555}
   \langle T^{\mu \nu} \rangle=\dfrac{1}{\pi}\bigg(\dfrac{2}{t^{2}}+\dfrac{\Phi^{2}(v)}{2}\bigg)\bigg(\dfrac{t^{\mu}t^{\nu}}{t^{2}}-\dfrac{1}{2}\eta^{\mu \nu}\bigg)+\dfrac{\Phi^{2}(v)}{2\pi}\bigg(\ln{\bigg[\dfrac{\Phi(v)\sqrt{-t^{2}}}{2}\bigg]}+\gamma\bigg)\eta^{\mu \nu}-\dfrac{1}{48\pi}\{a(v),v\}
   \left[ \begin{array}{cc}
   1& 1 \\
    1 & 1 \\
  \end{array}  \right] \ ,
\end{equation}
where $t^{2}=2t^{u}t^{v}$. The obtained expression explicitly depends on $t^\mu$ --- on the vector along which the point splitting is done. To restore the symmetry of the expressions under consideration we have to perform the averaging over all possible direction of $t^\mu$. The averaging is done in the standard manner using that:

\begin{equation}
    \label{e:34}
     \overline{t^{\mu} t^{\nu}} = \dfrac{1}{2}t^{2}\eta^{\mu \nu}.
\end{equation}
Then, for the expectation value under consideration we obtain the expression in eq. (\ref{e:35}), where $\sqrt{-t^2} \equiv \delta$.

\section{The change of the flux under the transformation (\ref{eq:cft5})} \label{G}

The $vv$ component of the stress energy tensor is given by (\ref{e:3}). Its regularised form (upto the $2\pi$ factor) we define as:

\begin{eqnarray}
    \label{ft2}
    T(v)\equiv -2\pi\dfrac{i}{2}\lim_{v^{\prime} \rightarrow v}\bigg[\bigg(\sqrt{2}\partial_{v}-\sqrt{2}\partial_{v^{\prime}}\bigg)\langle \psi^{\dagger}_{2}(v^{\prime},u)\psi_{2}(v,u)\rangle+\dfrac{i}{\pi}\dfrac{1}{(v-v^{\prime})^{2}}\bigg]= \nonumber \\
    =-i\pi\lim_{\delta \rightarrow 0}\bigg[\bigg(\sqrt{2}\partial_{v}-\sqrt{2}\partial_{v^{\prime}}\bigg)\langle \psi^{\dagger}_{2}(v^{\prime},u)\psi_{2}(v,u)\rangle\bigg|_{v^{\prime}=v-i\delta}-\dfrac{i}{\pi}\dfrac{1}{\delta^{2}}\bigg].
\end{eqnarray}
We want to find the transformation law of the \eqref{ft2} under the transformation (\ref{eq:cft5}).
First of all, note that we use point splitting regularization: $v^{\prime}=v-i\delta$. After transformation (\ref{eq:cft5}) we have that

\begin{equation}
    \label{ft4}
    \bar{v}^{\prime}-\bar{v} \equiv -is= \bar{v}(v-i\delta)-\bar{v}(v)=-i\dfrac{\Phi^{2}(v)}{m^{2}}\delta-\dfrac{\Phi(v) \Phi^{\prime}(v)}{m^{2}}\delta^{2}+\dfrac{i}{3}\bigg(\Phi^{\prime \prime}(v)\Phi(v)+\Phi^{\prime \ 2}(v)\bigg)\delta^{3}+... .
\end{equation}
Hence $T(v)$ transforms as

\begin{eqnarray}
    \label{ft5}
    T(v)=-i\pi\lim_{\delta \rightarrow 0}\bigg[\bigg(\sqrt{2}\partial_{v}-\sqrt{2}\partial_{v^{\prime}}\bigg)\langle \psi^{\dagger}_{2}(v^{\prime},u)\psi_{2}(v,u)\rangle\bigg|_{v^{\prime}=v-i\delta}-\dfrac{i}{\pi}\dfrac{1}{\delta^{2}}\bigg]= \nonumber \\
    =-i\pi \bigg[ \bigg(\dfrac{\Phi^{3}(v)\Phi(v^{\prime})}{m^{4}}\sqrt{2}\partial_{\bar{v}}-\dfrac{\Phi^{3}(v^{\prime})\Phi(v)}{m^{4}}\sqrt{2}\partial_{\bar{v}^{\prime}}\bigg)\langle \eta^{\dagger}_{2}\big(\bar{v}(v^{\prime}),\bar{u}\big)\eta_{2}\big(\bar{v}(v),\bar{u}\big)\rangle+ \nonumber \\
    +\dfrac{\sqrt{2}}{m^{2}}\bigg(\Phi^{\prime}(v)\Phi(v^{\prime})-\Phi(v)\Phi^{\prime}(v^{\prime})\bigg)\langle \eta^{\dagger}_{2}\big(\bar{v}(v^{\prime}),\bar{u}\big)\eta_{2}\big(\bar{v}(v),\bar{u}\big)\rangle -\dfrac{i}{\pi}\dfrac{1}{\delta^{2}} \bigg]= \nonumber \\
    =-i\pi\bigg[\bigg(\dfrac{d\bar{v}}{dv}\bigg)^{2}\bigg[\sqrt{2}\partial_{\bar{v}}-\sqrt{2}\partial_{\bar{v}^{\prime}}\bigg]\langle \eta^{\dagger}_{2}\big(\bar{v}(v^{\prime}),\bar{u}\big)\eta_{2}\big(\bar{v}(v),\bar{u}\big)\rangle+ \nonumber \\
    +\bigg(\dfrac{\Phi^{3}(v)\Phi(v^{\prime})-\Phi^{4}(v)}{m^{4}}\sqrt{2}\partial_{\bar{v}}-\dfrac{\Phi^{3}(v^{\prime})\Phi(v)-\Phi^{4}(v)}{m^{4}}\sqrt{2}\partial_{\bar{v}^{\prime}}\bigg)\langle \eta^{\dagger}_{2}\big(\bar{v}(v^{\prime}),\bar{u}\big)\eta_{2}\big(\bar{v}(v),\bar{u}\big)\rangle+\nonumber \\
    +\dfrac{\sqrt{2}}{m^{2}}\bigg(\Phi^{\prime}(v)\Phi(v^{\prime})-\Phi(v)\Phi^{\prime}(v^{\prime})\bigg)\langle \eta^{\dagger}_{2}\big(\bar{v}(v^{\prime}),\bar{u}\big)\eta_{2}\big(\bar{v}(v),\bar{u}\big)\rangle -\dfrac{i}{\pi}\dfrac{1}{\delta^{2}} \bigg]=\nonumber \\
    =\bigg(\dfrac{d\bar{v}}{dv}\bigg)^{2}\Tilde{T}(\bar{v})-i \pi \bigg[-\dfrac{i}{\pi}\dfrac{1}{\big(\bar{v}(v)-\bar{v}(v^{\prime})\big)^{2}}-\dfrac{i}{\pi}\dfrac{1}{\delta^{2}}+ \nonumber \\
    +\bigg(\dfrac{\Phi^{3}(v)\Phi(v^{\prime})-\Phi^{4}(v)}{m^{4}}\sqrt{2}\partial_{\bar{v}}-\dfrac{\Phi^{3}(v^{\prime})\Phi(v)-\Phi^{4}(v)}{m^{4}}\sqrt{2}\partial_{\bar{v}^{\prime}}\bigg)\langle \eta^{\dagger}_{2}\big(\bar{v}(v^{\prime}),\bar{u}\big)\eta_{2}\big(\bar{v}(v),\bar{u}\big)\rangle+\nonumber \\
    +\dfrac{\sqrt{2}}{m^{2}}\bigg(\Phi^{\prime}(v)\Phi(v^{\prime})-\Phi(v)\Phi^{\prime}(v^{\prime})\bigg)\langle \eta^{\dagger}_{2}\big(\bar{v}(v^{\prime}),\bar{u}\big)\eta_{2}\big(\bar{v}(v),\bar{u}\big)\rangle \bigg],
\end{eqnarray}
where
\begin{equation}
\label{ft7}
    \Tilde{T}(\bar{v})=-i \pi \lim_{\bar{v}^{\prime} \rightarrow \bar{v}} \bigg(\bigg[\sqrt{2}\partial_{\bar{v}}-\sqrt{2}\partial_{\bar{v}^{\prime}}\bigg]\langle \eta^{\dagger}_{2}\big(\bar{v}^{\prime},\bar{u}\big)\eta_{2}\big(\bar{v},\bar{u}\big)\rangle+\dfrac{i}{\pi}\dfrac{1}{\big(\bar{v}-\bar{v}^{\prime}\big)^{2}}\bigg).
\end{equation}
Using that in the limit $\delta \rightarrow 0$
\begin{equation}
    \label{ft8}
    \dfrac{\sqrt{2}}{m^{2}}\bigg(\Phi^{\prime}(v)\Phi(v^{\prime})-\Phi(v)\Phi^{\prime}(v^{\prime})\bigg)\langle \eta^{\dagger}_{2}\big(\bar{v}(v^{\prime}),\bar{u}\big)\eta_{2}\big(\bar{v}(v),\bar{u}\big)\rangle=\dfrac{i}{2\pi}\bigg(\dfrac{\Phi^{\prime \ 2}}{\Phi^{2}}+\dfrac{\Phi^{\prime \prime}}{\Phi}\bigg),
\end{equation}
and
\begin{eqnarray}
    \label{ft9}
    \bigg(\dfrac{\Phi^{3}(v)\Phi(v^{\prime})-\Phi^{4}(v)}{m^{4}}\sqrt{2}\partial_{\bar{v}}-\dfrac{\Phi^{3}(v^{\prime})\Phi(v)-\Phi^{4}(v)}{m^{4}}\sqrt{2}\partial_{\bar{v}^{\prime}}\bigg)\langle \eta^{\dagger}_{2}\big(\bar{v}(v^{\prime}),\bar{u}\big)\eta_{2}\big(\bar{v}(v),\bar{u}\big)\rangle= \nonumber \\
    =\dfrac{i}{2\pi}\bigg(5\dfrac{\Phi^{\prime \ 2}}{\Phi^{2}}-2\dfrac{\Phi^{\prime \prime}}{\Phi}-4i\dfrac{1}{\delta}\dfrac{\Phi^{\prime }}{\Phi}\bigg),
\end{eqnarray}
and
\begin{equation}
    \label{ft10}
    -\dfrac{i}{\pi}\dfrac{1}{\big(\bar{v}(v)-\bar{v}(v^{\prime})\big)^{2}}-\dfrac{i}{\pi}\dfrac{1}{\delta^{2}}=\dfrac{i}{\pi}\bigg(-\dfrac{7}{3}\dfrac{\Phi^{\prime \ 2}}{\Phi^{2}}+\dfrac{2}{3}\dfrac{\Phi^{\prime \prime}}{\Phi}+2i\dfrac{1}{\delta}\dfrac{\Phi^{\prime }}{\Phi}\bigg)
\end{equation}
we find that
\begin{eqnarray}
    \label{ft11}
    T(v)=\bigg(\dfrac{d\bar{v}}{dv}\bigg)^{2}\Tilde{T}(\bar{v})+\dfrac{1}{3}\bigg(\dfrac{1}{2}\dfrac{\Phi^{\prime \prime}}{\Phi}-\dfrac{\Phi^{\prime \ 2}}{\Phi^{2}}\bigg)=\bigg(\dfrac{d\bar{v}}{dv}\bigg)^{2}\bigg[\Tilde{T}(\bar{v})-\dfrac{1}{12}\{v(\bar{v}),\bar{v}\}\bigg]= \nonumber \\
    =\bigg(\dfrac{dv}{d\bar{v}}\bigg)^{-2}\bigg[\Tilde{T}(\bar{v})-\dfrac{1}{12}\{v(\bar{v}),\bar{v}\}\bigg],
\end{eqnarray}
i.e. $T_{vv}$ indeed transforms affinely by a shift by a Schwarzian derivative.

%\newpage

\end{document}